\documentclass{aastex62}
\usepackage[colorinlistoftodos]{todonotes}
\usepackage{CJK}
\submitjournal{ApJS}
\shorttitle{Exoplanets in the Antarctic sky. II.}
\shortauthors{Zhang et al.}
\begin{document}

\begin{CJK*}{UTF8}{gbsn}
\title{Exoplanets in the Antarctic sky. II. 116 Transiting Exoplanet Candidates Found by AST3-II within the Southern CVZ of \textit{TESS}}

\correspondingauthor{Hui Zhang}
\email{huizhang@nju.edu.cn}
\correspondingauthor{Ji-lin Zhou}
\email{zhoujl@nju.edu.cn}

\author{Hui Zhang (张辉)}
\affil{School of Astronomy and Space Science, Key Laboratory of Modern Astronomy and Astrophysics in Ministry of Education, \\
Nanjing University, Nanjing 210023, Jiangsu,China}

\author{Zhouyi Yu}
\affil{School of Astronomy and Space Science, Key Laboratory of Modern Astronomy and Astrophysics in Ministry of Education, \\
Nanjing University, Nanjing 210023, Jiangsu,China}

\author{Ensi Liang}
\affil{School of Astronomy and Space Science, Key Laboratory of Modern Astronomy and Astrophysics in Ministry of Education, \\
Nanjing University, Nanjing 210023, Jiangsu,China}

\author{Ming Yang}
\affil{School of Astronomy and Space Science, Key Laboratory of Modern Astronomy and Astrophysics in Ministry of Education, \\
Nanjing University, Nanjing 210023, Jiangsu,China}

\author{Michael C. B. Ashley}
\affil{School of Physics, University of New South Wales, NSW 2052, Australia}

\author{Xiangqun Cui}
\affil{Nanjing Institute of Astronomical Optics and Technology, Nanjing 210042, China}
\affil{Chinese Center for Antarctic Astronomy, Nanjing 210008, China}

\author{Fujia Du}
\affil{Nanjing Institute of Astronomical Optics and Technology, Nanjing 210042, China}
\affil{Chinese Center for Antarctic Astronomy, Nanjing 210008, China}

\author{Jianning Fu}
\affil{Department of Astronomy, Beijing Normal University, Beijing, 100875, China}

\author{Xuefei Gong}
\affil{Nanjing Institute of Astronomical Optics and Technology, Nanjing 210042, China}
\affil{Chinese Center for Antarctic Astronomy, Nanjing 210008, China}

\author{Bozhong Gu}
\affil{Nanjing Institute of Astronomical Optics and Technology, Nanjing 210042, China}
\affil{Chinese Center for Antarctic Astronomy, Nanjing 210008, China}

\author{Yi Hu}
\affil{National Astronomical Observatories, Chinese Academy of Sciences, Beijing 100012, China}
\affil{Chinese Center for Antarctic Astronomy, Nanjing 210008, China}

\author{Peng Jiang}
\affil{Polar Research Institute of China, Shanghai 200136, China}
\affil{Chinese Center for Antarctic Astronomy, 451 Jinqiao Rd, Nanjing 210008, China}

\author{Huigen Liu}
\affil{School of Astronomy and Space Science, Key Laboratory of Modern Astronomy and Astrophysics in Ministry of Education, \\
Nanjing University, Nanjing 210023, Jiangsu,China}

\author{Jon Lawrence}
\affil{Australian Astronomical Optics, Macquarie University, NSW 2109, Australia}

\author{Qiang Liu}
\affil{National Astronomical Observatories, Chinese Academy of Sciences, Beijing 100012, China}

\author{Xiaoyan Li}
\affil{Nanjing Institute of Astronomical Optics and Technology, Nanjing 210042, China}
\affil{Chinese Center for Antarctic Astronomy, Nanjing 210008, China}

\author{Zhengyang Li}
\affil{Nanjing Institute of Astronomical Optics and Technology, Nanjing 210042, China}
\affil{Chinese Center for Antarctic Astronomy, Nanjing 210008, China}

\author{Bin Ma}
\affil{National Astronomical Observatories, Chinese Academy of Sciences, Beijing 100012, China}
\affil{Chinese Center for Antarctic Astronomy, Nanjing 210008, China}
\affil{University of Chinese Academy of Sciences, Beijing 100049, China}

\author{Jeremy Mould}
\affil{Centre for Astrophysics and Supercomputing, Swinburne University of Technology, PO Box 218, Mail Number H29, Hawthorn, VIC 3122, Australia}
\affil{ARC Centre of Excellence for All-sky Astrophysics (CAASTRO)}

\author{Zhaohui Shang}
\affil{Tianjin Astrophysics Center, Tianjin Normal University, Tianjin 300387, China}
\affil{National Astronomical Observatories, Chinese Academy of Sciences, Beijing 100012, China}
\affil{Chinese Center for Antarctic Astronomy, Nanjing 210008, China}

\author{Nicholas B. Suntzeff}
\affil{George P. and Cynthia Woods Mitchell Institute for Fundamental Physics \& Astronomy, Texas A. \& M. University, \\
Department of Physics and Astronomy, 4242 TAMU, College Station, TX 77843, US}

\author{Charling Tao}
\affil{Aix Marseille Univ, CNRS/IN2P3, CPPM, Marseille, France}
\affil{Physics Department and Tsinghua Center for Astrophysics (THCA), Tsinghua University, Beijing, 100084, China}

\author{Qiguo Tian}
\affil{Polar Research Institute of China, 451 Jinqiao Rd, Shanghai 200136, China}

\author{C. G. Tinney}
\affil{Exoplanetary Science at UNSW, School of Physics, UNSW Sydney, NSW 2052, Australia}

\author{Syed A. Uddin}
\affil{Purple Mountain Observatory, Nanjing 210008, China}

\author{Lifan Wang}
\affil{George P. and Cynthia Woods Mitchell Institute for Fundamental Physics \& Astronomy, Texas A. \& M. University, \\
Department of Physics and Astronomy, 4242 TAMU, College Station, TX 77843, US}
\affil{Purple Mountain Observatory, Nanjing 210008, China}
\affil{Chinese Center for Antarctic Astronomy, Nanjing 210008, China}

\author{Songhu Wang}
\affil{Department of Astronomy, Yale University, New Haven, CT 06511, USA}

\author{Xiaofeng Wang}
\affil{Physics Department and Tsinghua Center for Astrophysics (THCA), Tsinghua University, Beijing, 100084, China}

\author{Peng Wei}
\affil{School of Astronomy and Space Science, Key Laboratory of Modern Astronomy and Astrophysics in Ministry of Education, \\
Nanjing University, Nanjing 210023, Jiangsu,China}

\author{Duncan Wright}
\affil{University of Southern Queensland, Computational Engineering and Science Research Centre, Toowoomba, Queensland 4350, Australia}

\author{Xuefeng Wu}
\affil{Purple Mountain Observatory, Nanjing 210008, China}
\affil{Chinese Center for Antarctic Astronomy, Nanjing 210008, China}

\author{Robert A. Wittenmyer}
\affil{University of Southern Queensland, Computational Engineering and Science Research Centre, Toowoomba, Queensland 4350, Australia}

\author{Lingzhe Xu}
\affil{Nanjing Institute of Astronomical Optics and Technology, Nanjing 210042, China}

\author{Shi-hai Yang}
\affil{Nanjing Institute of Astronomical Optics and Technology, Nanjing 210042, China}
\affil{Chinese Center for Antarctic Astronomy, Nanjing 210008, China}

\author{Ce Yu}
\affil{School of Computer Science and Technology, Tianjin University, Tianjin 300072, China}

\author{Xiangyan Yuan}
\affil{Nanjing Institute of Astronomical Optics and Technology, Nanjing 210042, China}
\affil{Chinese Center for Antarctic Astronomy, Nanjing 210008, China}

\author{Jessica Zheng}
\affil{Australian Astronomical Observatory, 105 Delhi Road, North Ryde, NSW 2113, Australia}

\author{Hongyan Zhou}
\affil{Polar Research Institute of China, 451 Jinqiao Rd, Shanghai 200136, China}

\author{Ji-lin Zhou}
\affil{School of Astronomy and Space Science, Key Laboratory of Modern Astronomy and Astrophysics in Ministry of Education, \\
Nanjing University, Nanjing 210023, Jiangsu,China}

\author{Zhenxi Zhu}
\affil{Purple Mountain Observatory, Nanjing 210008, China}
\affil{Chinese Center for Antarctic Astronomy, Nanjing 210008, China}



\begin{abstract}
We report first results from the CHinese Exoplanet Searching Program from Antarctica (CHESPA)---a wide-field high-resolution photometric survey for transiting exoplanets carried out using telescopes of the AST3 (Antarctic Survey Telescopes times 3) project. There are now three telescopes (AST3-I, AST3-II, and CSTAR-II) operating at Dome A---the highest point on the Antarctic Plateau---in a fully automatic and remote mode to exploit the superb observing conditions of the site, and its long and uninterrupted polar nights. The search for transiting exoplanets is one of the key projects for AST3. During the austral winters of 2016 and 2017 we used the AST3-II telescope to survey a set of target fields near the southern ecliptic pole, falling within the continuous viewing zone of the \textit{TESS} mission \citep{Ricker10}. The first data release of the 2016 data, including images, catalogs and lightcurves of 26578 bright stars ($7.5\le \textbf{\textit{m}}_\textit{i} \le15$) was presented in \citet{Zhang18}. The best precision, as measured by the RMS of the lightcurves at the optimum magnitude of the survey ($\textbf{\textit{m}}_\textit{i}=10$), is around 2\,mmag. We detect 222 objects with plausible transit signals from these data, 116 of which are plausible transiting exoplanet candidates according to their stellar properties as given by the \textit{TESS} Input Catalog \citep{Stassun17}, Gaia DR2 \citep{Gaia18} and \textit{TESS}-\textit{HERMES} spectroscopy \citep{Sharma18}. With the first data release from \textit{\textit{TESS}} expected in late 2018, this candidate list will be a timely for improving the rejection of potential false positives.

\end{abstract}



\section{Introduction} \label{sec:intro}
The rapidly expanding sample size of exoplanets discovered over the last two decades has allowed entirely new classes of study of planetary demographics, and has dramatically expanded our understanding of the distribution of planetary orbital parameters. To further increase this understanding, more exoplanet detections covering a wider parameter space are required---and in particular more detections of planets orbiting host stars bright enough for ground-based follow-up to measure dynamical masses. The \textit{TESS} mission's satellite \citep{Ricker10} was launched successfully in April 2018, and is expected to produce a substantial crop of exactly this class of exoplanets \citep{Stassun17}, allowing it to have potentially even larger impact than the Kepler \citep{Borucki10} and CoRoT \citep{Auvergne09} missions.

In advance of these substantial detections of exoplanets by space telescopes, the first transiting exoplanets were discovered by ground-based, small-aperture telescopes, and these facilities have continued to operate in parallel over the last decade, delivering hundreds of exoplanets from projects including HATNET \citep{Bakos04}, WASP/SuperWASP \citep{Pollacco06}, HATSouth \citep{Bakos13}, and KELT \citep{Pepper07}. The common features of these ground-based searching programs have been the use of Small-apertures, Wide-fields, and Arrays of Telescopes (Hereafter the SWAT). And the SWAT have been proved to be one of the most efficient and economical ways to search for new transiting exoplanets from the ground.


However, photometric surveys using ground-based SWATs suffer from two major drawbacks compared to space-based surveys: lower photometric precision and lower duty-cycle coverage. As present and future space-based wide-field surveys continue to progress, from Kepler and CoRoT to \textit{TESS} to PLATO2.0 \citep{Barban13,Rauer14}, these drawbacks have called the value of the ground-based SWAT facilities into question. Ground-based SWAT surveys have to improve their capabilities and work in partnership with present and future space-based wide-field surveys to achieve the greatest impact. On one hand, as experience has been gained and new technologies adopted, new generation wide-field transit searching programs such as NGTS \citep{Wheatley18} and Pan-Planets \citep{Obermeier16} have been pushing photometric precision to levels precision (i.e., a few millimagnitudes). Their experience has shown that instruments that are stable over the long-term, combined with optimized operations (e.g., precision auto-guiding) are critical for improving long-term photometric precision. On the other hand, choosing a good site with superb observing conditions (e.g., a steady atmosphere and a clear dark sky) is also essential to guaranteeing an efficient observing by a ground-based SWAT facility.

Because of its extremely cold, dry, and clear atmosphere, the Antarctic Plateau provides favorable conditions for optical, infrared and THz astronomical observations. \citep{Lawrence04} reported a median seeing of $0.23^{\prime\prime}$ (average of $0.27^{\prime\prime}$) above a 30 m boundary layer at Dome C, drawing world-wide attention. Subsequently, many studies have been focused on the astronomical conditions at a variety of Antarctic sites, and have shown low sky brightness and extinction \citep{Kenyon06,Zou10,Yang17}, low water vapor \citep{Shi16}, very low wind speeds, and exceptional seeing above a thin boundary layer \citep{Bonner10,Okita13,Hu14,Aristidi09,Fossat10,Giordano12}. Furthermore, the decreased high-altitude turbulence above the plateau results in reduced scintillation noise, further improving photometric precision \citep{Kenyon06}. \cite{Saunders09} studied eight major factors, such as the boundary layer thickness, cloud coverage, auroral emission, airglow, atmospheric thermal backgrounds, precipitable water vapor, telescope thermal backgrounds, and the free-atmosphere seeing, at Domes A, B, C and F, and also Ridge A and B. After a systematic comparison, they concluded that Dome A, the highest point of the Antarctic Plateau, would be the best site overall.

In addition to the excellent photometric conditions at Dome A, the small variation in the elevation of targets as they track around the sky at Dome A will reduce the systematics. And most importantly, the uninterrupted polar nights offer an opportunity to obtain nearly continuous photometric monitoring for periods of more than a month. As shown by a series of previous studies \citep{Crouzet10,Daban10,Law13}, this greatly increases the detectability of transiting exoplanets with orbital periods longer than a few days. The outstanding photometric advantages of the Antarctic Plateau have been shown by the observing facilities at different sites, such as SPOT \citep{Taylor88} at the South Pole, the small-IRAIT \citep{Tosti06}, ASTEP-South \citep{Crouzet10} and ASTEP-400 \citep{Daban10, Mekarnia16} at Dome C, and CSTAR \citep{Wang11,Wang14,Yang15,Zong15,Oelkers16,Liang16} and AST3-I \citep{Wang17,Ma18} at Dome A.

To utilize the superb observing conditions, the construction of a remote observatory at Dome A commenced in 2008, with the installation of a first generation telescope CSTAR (the Chinese Small Telescope ARray, \citealp{Yuan08,Zhou10}). Based on a successful experience with CSTAR (and the lessons learned from it), a second generation of survey telescopes---the AST3 telescopes (Antarctic Survey Telescopes times 3, \citealp{Cui08,Yuan14})---were conceived to implement wide-field high-resolution photometric surveys at Dome A. The first and second AST3 telescopes---AST3-I and AST3-II---were installed at Dome A in 2012 and 2015 by the 28th and 29th CHINARE (CHInese National Antarctic Research Expeditions), respectively. Using the CSTAR and AST3 telescopes, we have been running an exoplanet survey program called CHESPA (the CHinese Exoplanet Searching Program from Antarctica). The primary science goal of CHESPA is finding Super-Earth or Neptune-sized transiting exoplanets around a variety of host-star stellar types that are sufficiently bright for radial velocity confirmation and dynamical mass measurement. The combination of dynamical masses and planetary sizes will allow us to determine the true masses, orbital eccentricities, and most critically their bulk densities. With accurate physical and dynamic properties determined, these exoplanets will be good targets for a wide range of characterization techniques, e.g., studies of their atmospheric structure and composition (see, e.g., \citealp{Seager07,Baraffe08}). The first six exoplanet candidates around the South Celestial Pole were reported by \cite{Wang14}, and (although not yet confirmed) they demonstrate the potential power of CHESPA for the discovery of new exoplanets.

Although the basic design of CHESPA is much the same as most other ground-based SWAT facilities, we do have a few additional advantages over other, similar surveys:
a large FoV combined with a relatively high angular resolution of $1^{\prime\prime}$ pixel$^{-1}$, multi-band filters, and superb observing conditions from Antarctica. To maximize our linkage with upcoming space-based surveys like \textit{TESS}, we have scheduled our ongoing survey to make observations that scan \textit{TESS}' high-value target zones, but with a higher angular resolution and in a different filter (Sloan $i$-band). During the austral winters of 2016 and 2017, we have surveyed two sets of fields selected within the Southern Continuous Viewing Zone (CVZ) of \textit{TESS} \citep{Ricker10} using  AST3-II. The remaining target fields will be scanned in 2019. The reason we choose these fields is that the stars in the CVZ will be over a 13 times longer observing period than most of objects of the \textit{TESS} survey (i.e. continuously over a full year). Hence, light curves for objects in the CVZ will be much more sensitive to small planets in short-period orbits \citep{Stassun17}, as well as sensitive to larger planets with orbits with much longer periods. In addition, the \textit{TESS} CVZ and the JWST CVZ are both over the same area of sky (i.e. the south ecliptic pole), so planets detected in these overlapping regions will have enormous potential for further detailed follow-up to characterize in detail their atmosphere and internal structure.

\cite{Zhang18} has presented the first release of 2016 data from AST3-II, as well as a catalog of 221 newly discovered variables. In this paper, we present the detection of 116 transiting exoplanet candidates from the same data. As more data are returned from Antarctica, new results will be presented in forthcoming papers. In this work, we introduce the AST3-II facility in Section \ref{sec:instru} and describe the observations of CHESPA briefly in Section \ref{sec:survey}; in Section \ref{sec:pipeline} we detail the data reduction pipeline, including lightcurve detrending, transit signal searching, and the transit signal validation software modules; finally, we present the survey results in Section \ref{sec:results} and summarize the paper in Section \ref{sec:summary}.

\section{Instruments} \label{sec:instru}
Detailed descriptions of the AST3 telescopes have been previously presented by \cite{Cui08,Yuan14,Yuan15,Wang17}. Here we only focus on the key properties of the AST3-II telescope. The AST3-II telescope (which acquired all the data presented in this paper) has an effective aperture of 50 cm. It is designed to obtain wide-field ($1.5 \times 2.9 \mbox{ deg}^2$ in RA $\times$ Dec) and high resolution ($\approx 1^{\prime\prime}$ pixel$^{-1}$) imaging in the Sloan $i$-band. It employs a modified Schmidt optical design \citep{Yuan12} and a $10\mbox{K} \times10\mbox{K}$\,pixel frame transfer STA1600FT CCD camera. To withstand the extremely harsh environment at Dome A, careful design work has been done to implement multiple innovations in the telescope's snow-proofing and defrosting systems. As a result, AST3-II worked well during the extremely cold austral winters of 2016 and 2017, and acquired over 30\,TB of high-quality images. During the observational seasons in austral winter, AST3-II is operated remotely and the scheduled observations are executed in a fully-automatic mode. The hardware and software for the facility---including the hardware operation monitor, telescope control computer, and data storage array---were developed by the National Astronomical Observatories, Chinese Academy of Sciences (NAOC) \citep{Shang12,Hu16}. The electrical power supply and internet communication were provided by a similarly reliable on-site observatory platform, PLATO-A, which is an improved version of the PLATO system developed by UNSW Sydney as an automated observatory platform for CSTAR and other earlier instruments. PLATO-A was designed to provide a continuous 1\,kW power source, a warm environment for equipment, and internet communications to the AST3 telescopes for a year without servicing \citep{Lawrence09,Ashley10}.

\section{Observations} \label{sec:survey}
\begin{figure}
\centering
\plotone{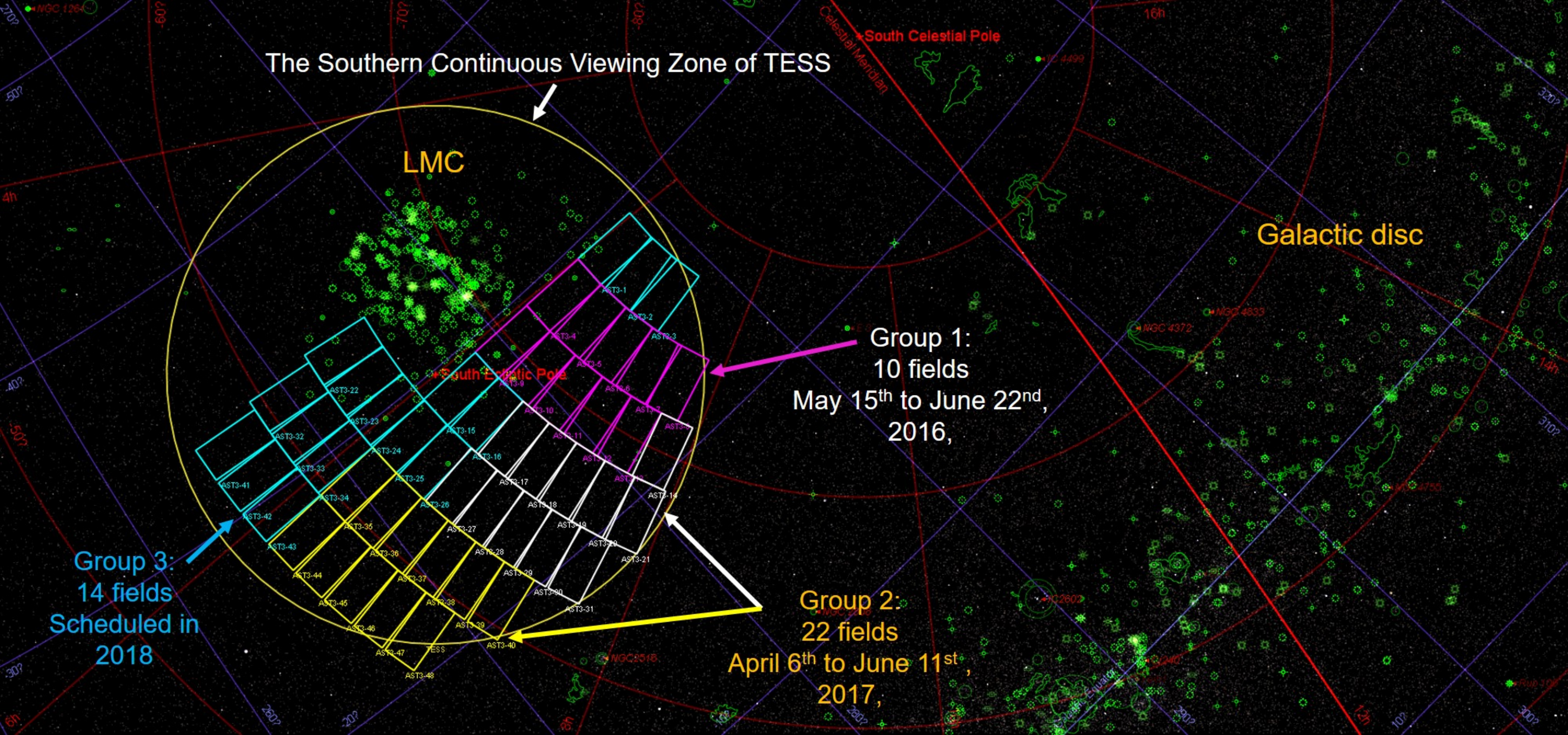}
\caption{Scheduled survey of 48 target fields in 2016, 2017, and 2018. Each field has a sky coverage of $\sim4.3$ deg$^2$. Fields close to the LMC are excluded to avoid crowded fields of giant stars. Group 1 (10 fields) and group 2 (22 fields) have been scanned in 2016 and 2017, respectively.\label{fig:48fields}}
\end{figure}

The scientific background and observational strategy of CHESPA are described in \cite{Zhang18}. We summarize the key points here. The CHESPA program is dedicated to searching for transiting exoplanets in the southern polar sky at highly negative declinations. It has been running since 2012 using the CSTAR and AST3 telescopes, with a first batch of six exoplanet candidates published by \cite{Wang14}. To maximize collaboration with \textit{TESS} and enhance the scientific importance of our searching program, we selected 48 target fields close to the South Ecliptic Pole ($\mbox{RA} =06^{\mbox{h}}\,00^{\mbox{m}}\,00^{\mbox{s}} , \mbox{Dec} =-66^\circ\,33^{\prime}\,00^{\prime\prime}$) and within the \textit{TESS}' Southern CVZ (Continuous Viewing Zone) (see Figure \ref{fig:48fields} and Table \ref{tab:48fields} for details). All target fields are and are suitable for low-airmass observation from Antarctica. Target stars located in these fields will also be monitored by \textit{TESS} for continuous 12 month period. Thus, any candidate of interest found within these fields may be followed up and studied thoroughly in the future.

\begin{deluxetable}{c|c|c|l|r}
\tablecaption{Center coordinates of the 48 AST3-II target fields surveyed/scheduled in 2016--2019. \label{tab:48fields}}
\tablehead{
\colhead{Field Name} & \multicolumn{2}{c}{Field Center} & \colhead{Observation Date} &\colhead{Valid Images} \\
\colhead{} & \colhead{RA(J2000)} & \colhead{Dec(J2000)} &\colhead{} &\colhead{}
}
\startdata
AST3II001 & $ 93.000  $ & $ -76.000 $  & 2018--2019               &             \\
AST3II002 & $ 99.000  $ & $ -76.000 $  & 2018--2019               &             \\
AST3II003 & $ 105.000 $ & $ -76.000 $  & 2018--2019               &             \\
AST3II004 & $ 93.000  $ & $ -73.000 $  & 2016.05.15--2016.06.22   &  3179       \\
AST3II005 & $ 98.500  $ & $ -73.000 $  & 2016.05.15--2016.06.22   &  3080       \\
AST3II006 & $ 104.000 $ & $ -73.000 $  & 2016.05.15--2016.06.22   &  3049       \\
AST3II007 & $ 109.500 $ & $ -73.000 $  & 2016.05.15--2016.06.22   &  3103       \\
AST3II008 & $ 115.000 $ & $ -73.000 $  & 2016.05.15--2016.06.22   &  3248       \\
AST3II009 & $ 93.000  $ & $ -70.000 $  & 2016.05.15--2016.06.22   &  3090       \\
AST3II010 & $ 97.750  $ & $ -70.000 $  & 2016.05.15--2016.06.22   &  3000       \\
AST3II011 & $ 102.500 $ & $ -70.000 $  & 2016.05.15--2016.06.22   &  2991       \\
AST3II012 & $ 107.250 $ & $ -70.000 $  & 2016.05.15--2016.06.22   &  3021       \\
AST3II013 & $ 112.000 $ & $ -70.000 $  & 2016.05.15--2016.06.22   &  3128       \\
AST3II014 & $ 116.750 $ & $ -70.000 $  & 2017.04.06--2017.05.12   &  $\ge3000$  \\
AST3II015 & $ 93.000  $ & $ -67.000 $  & 2018--2019               &             \\
AST3II016 & $ 97.000  $ & $ -67.000 $  & 2018--2019               &             \\
AST3II017 & $ 101.000 $ & $ -67.000 $  & 2017.04.06--2017.05.12   &  $\ge3000$  \\
AST3II018 & $ 105.000 $ & $ -67.000 $  & 2017.04.06--2017.05.12   &  $\ge3000$  \\
AST3II019 & $ 109.000 $ & $ -67.000 $  & 2017.04.06--2017.05.12   &  $\ge3000$  \\
AST3II020 & $ 113.000 $ & $ -67.000 $  & 2017.04.06--2017.05.12   &  $\ge3000$  \\
AST3II021 & $ 117.000 $ & $ -67.000 $  & 2017.04.06--2017.05.12   &  $\ge3000$  \\
AST3II022 & $ 82.500  $ & $ -64.000 $  & 2018--2019               &             \\
AST3II023 & $ 86.000  $ & $ -64.000 $  & 2018--2019               &             \\
AST3II024 & $ 89.500  $ & $ -64.000 $  & 2018--2019               &             \\
AST3II025 & $ 93.000  $ & $ -64.000 $  & 2018--2019               &             \\
AST3II026 & $ 96.500  $ & $ -64.000 $  & 2018--2019               &             \\
AST3II027 & $ 100.000 $ & $ -64.000 $  & 2017.04.06--2017.05.12   &  $\ge3000$ \\
AST3II028 & $ 103.500 $ & $ -64.000 $  & 2017.04.06--2017.05.12   &  $\ge3000$  \\
AST3II029 & $ 107.000 $ & $ -64.000 $  & 2017.04.06--2017.05.12   &  $\ge3000$\\
AST3II030 & $ 110.500 $ & $ -64.000 $  & 2017.04.06--2017.05.12   &  $\ge3000$  \\
AST3II031 & $ 114.000 $ & $ -64.000 $  & 2017.04.06--2017.05.12   &  $\ge3000$  \\
AST3II032 & $ 83.250  $ & $ -61.000 $  & 2018--2019               &             \\
AST3II033 & $ 86.500  $ & $ -61.000 $  & 2018--2019               &             \\
AST3II034 & $ 89.750  $ & $ -61.000 $  & 2018--2019               &             \\
AST3II035 & $ 93.000  $ & $ -61.000 $  & 2017.05.14--2017.06.11   &  $\ge4100$ \\
AST3II036 & $ 96.250  $ & $ -61.000 $  & 2017.05.14--2017.06.11   &  $\ge4100$  \\
AST3II037 & $ 99.500  $ & $ -61.000 $  & 2017.05.14--2017.06.11   &  $\ge4100$  \\
AST3II038 & $ 102.750 $ & $ -61.000 $  & 2017.05.14--2017.06.11   &  $\ge4100$  \\
AST3II039 & $ 106.000 $ & $ -61.000 $  & 2017.05.14--2017.06.11   &  $\ge4100$ \\
AST3II040 & $ 109.250 $ & $ -61.000 $  & 2017.05.14--2017.06.11   &  $\ge4100$  \\
AST3II041 & $ 84.000  $ & $ -58.000 $  & 2018--2019               &             \\
AST3II042 & $ 87.000  $ & $ -58.000 $  & 2018--2019               &             \\
AST3II043 & $ 90.000  $ & $ -58.000 $  & 2018--2019               &             \\
AST3II044 & $ 93.000  $ & $ -58.000 $  & 2017.05.14--2017.06.11   &  $\ge4100$ \\
AST3II045 & $ 96.000  $ & $ -58.000 $  & 2017.05.14--2017.06.11   &  $\ge4100$  \\
AST3II046 & $ 99.000  $ & $ -58.000 $  & 2017.05.14--2017.06.11   &  $\ge4100$  \\
AST3II047 & $ 102.000 $ & $ -58.000 $  & 2017.05.14--2017.06.11   &  $\ge4100$  \\
AST3II048 & $ 105.000 $ & $ -58.000 $  & 2017.05.14--2017.06.11   &  $\ge4100$  \\
\enddata
\end{deluxetable}

These chosen fields are divided into three groups, each of which was originally scheduled for observation over an austral winter in 2016, 2017, and 2018. From 2016 May 16 to 2016 June 22 in 2016, we observed the first group consisting of 10 fields (AST3II004--AST3II013). The second group comprises 22 fields and observed from 2017 April 6 to 2017 June 11 (see Table \ref{tab:48fields} for details). The ten 2016 fields were monitored by the AST3-II telescope for over 350 hours, spanning 37 available or half-available nights. (The remaining time was allocated to instrument maintenance and used for other key projects, including a supernova search.) To avoid saturating bright stars, and to maximize the survey's dynamic range, we adopted a short-exposure-stacking strategy. The 10 target fields were scanned one by one in a loop with three consecutive 10\,s exposures being taken in each field, before moving on to the next field. The resulting sampling cadence is about 12 minutes for each field, including the dead time caused by slewing ($\sim 24$\,s) and CCD readout ($\sim48$\,s). Twilight sky frames were taken at each dawn and dusk during the periods when the Sun was still rising and setting from Dome A. These frames were then median-combined to produce a master flat-field image. To reduce the systematic errors caused by the imperfection of the flat-field correction, we adjusted the focus of the optics to sample stars with point-spread function sizes between $\mbox{FWHM}=3$ pixels to $\mbox{FWHM}=5$ pixels, while the pixel-scale of AST3-II is $1^{\prime\prime}$ pixel$^{-1}$, which is designed to match the average seeing at Dome A within the boundary layer. We made template images for each target field, and determined accurate astrometric solutions. Every time the telescope points to a new field, any small offsets in RA and Dec are corrected by cross-matching the first image with the template, and correcting pointing for the second and third images. \textbf{Then the last two frames are resampled and matched to the first one to guarantee pixel alignments. At last, all three pixel-aligned images are median combined by the \emph{Swarp} code \citep{Bertin02} to produce a new image}. 

During 2016, over 35000 science images in our 10 target fields were acquired by AST3-II and brought back on hard disks by the 33rd CHINARE team. The first data release, including 18729 coadded images/catalogs and 26578 lightcurves of stars brighter than 15th magnitude in the Sloan $i$-band were presented in \cite{Zhang18}. In 2017, more than 80000 images were taken and we await the return of this data from the next polar servicing expedition. \textbf{In 2018, however, due to some technical problems, no expedition was sent to the Kunlun station. And the scheduled CHESPA 2018 survey was canceled after the on-site fuel storage was exhausted. }

\subsection{Detection Probability versus Orbital Period}
\begin{figure}
\centering
\plotone{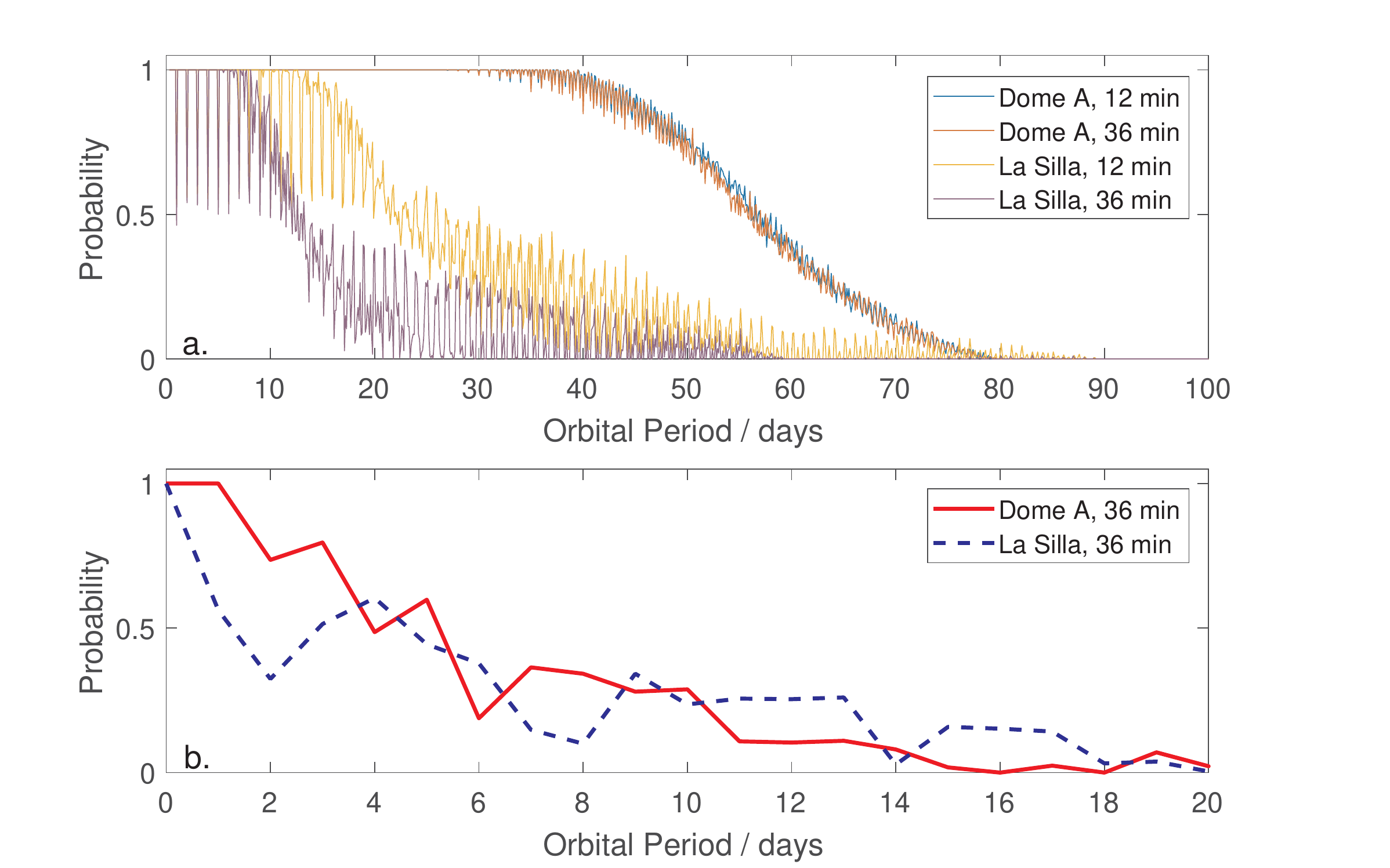}
\caption{Panel a. Simulated detection probabilities of transiting exoplanets observed from Dome A and La Silla, assuming the observations last from April to September; Panel b. Simulated detection probabilities of transiting exoplanets observed from Dome A and La Silla, using the real time series obtained in 2016.\label{fig:detection-prob}}
\end{figure}

\begin{figure}
\centering
\plotone{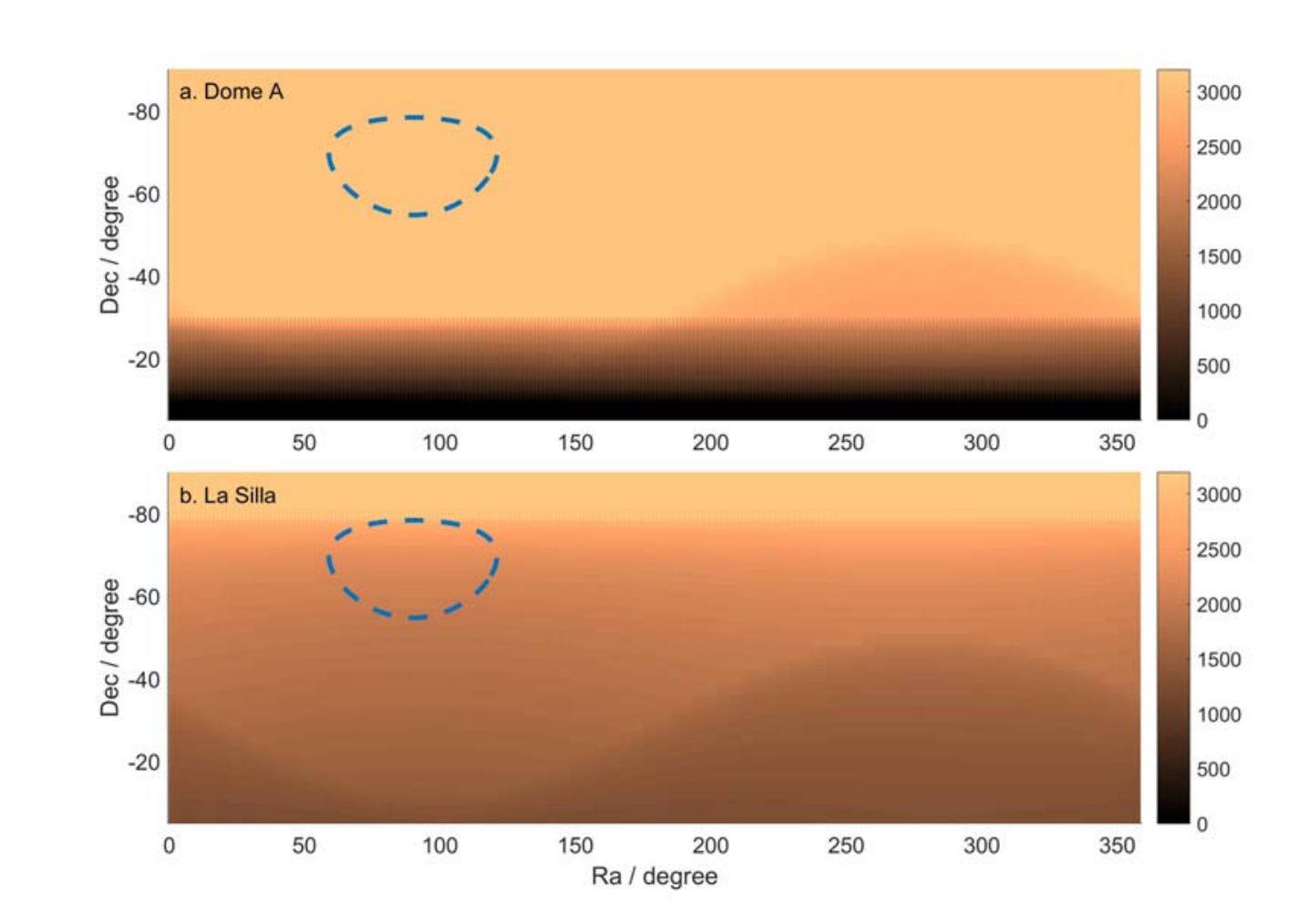}
\caption{Panel a. Total usable hours from Dome A, assuming the observations last from April to September; Panel b. Total usable hours from La Silla, assuming the observations last from April to September. The dashed circle denotes the Southern CVZ of \textit{TESS} in each panel. \label{fig:valid-hours}}
\end{figure}

To demonstrate the advantages of Antarctica for finding transiting exoplanets within the Southern CVZ of \textit{TESS}, we simulated the relation between the orbital period of a transiting exoplanet and its probability of being found at two sites: Dome A and La Silla (i.e. a representative temperate-latitude observing site with good weather and seeing). The simulated observation campaign lasts from the beginning of April to the end of September, which covers the entire austral winter from Dome A. We assume the fraction of bad weather to be $20\%$ at both sites. Two sampling cadences are adopted---12 and 36 minutes---consistent with our Dome A observing strategies. In addition, we performed simulations implementing the actual real-time epochs of observation from our 2016 data. \textbf{Due to some technical failures caused by the harsh environment at Dome A, the observation in 2016 ended in the end of June and the overall operation coverage is around $\sim 40\%$ which is much less than the coverage we expected: $\sim 75\%$. As a result, the detection probability in 2016 is not quite improved and it is even worse at the long-period end.} The results are shown in Figure \ref{fig:detection-prob}. The detection probability of a transiting planet is calculated following the method of \cite{Beatty08}. These figures highlight three key features. First, that the limited amount of actual data obtained in 2016 does not have significantly different sensitivity from a temperature latitude site. {\em But}, that more extended continuous observations with no diurnal gaps (i.e. from Dome A) {\em massively} improves the efficiency of the detection of planets at periods of longer than $\sim$10\,d. And finally, that when observing at Dome A the choice of a 12 minute or 36 minute cadence makes little difference to the detectability of planets out to orbital periods of $\sim$40\,d.
Figure \ref{fig:valid-hours} shows the total available hours as a function of position on the sky as visible from Dome A and La Silla. The dashed circle shows the position of the Southern CVZ of \textit{TESS}, which can clearly be monitored for a much longer period from Dome A, making the detection of planets at periods of 30-40\,d feasible from Dome A.


\section{Transit Signal Searching Pipeline} \label{sec:pipeline}
A detailed description of the ``Image Reduction Module''  is provided in \cite{Zhang18}, including the standard image processing steps (e.g., the 2-D over-scan subtraction, flat-field correction, zero-point magnitude correction), plus some treatments unique to our data (e.g., cross-talk correction and electromagnetic interference noise fringe correction). In this paper we focus on the lightcurve production process which comprises ``Lightcurve Detrending'', ``Transit Signal Searching'', and ``Transit Signal Validation'' modules.

\subsection{Lightcurve Detrending Module}

\begin{figure}
\includegraphics[width=\textwidth]{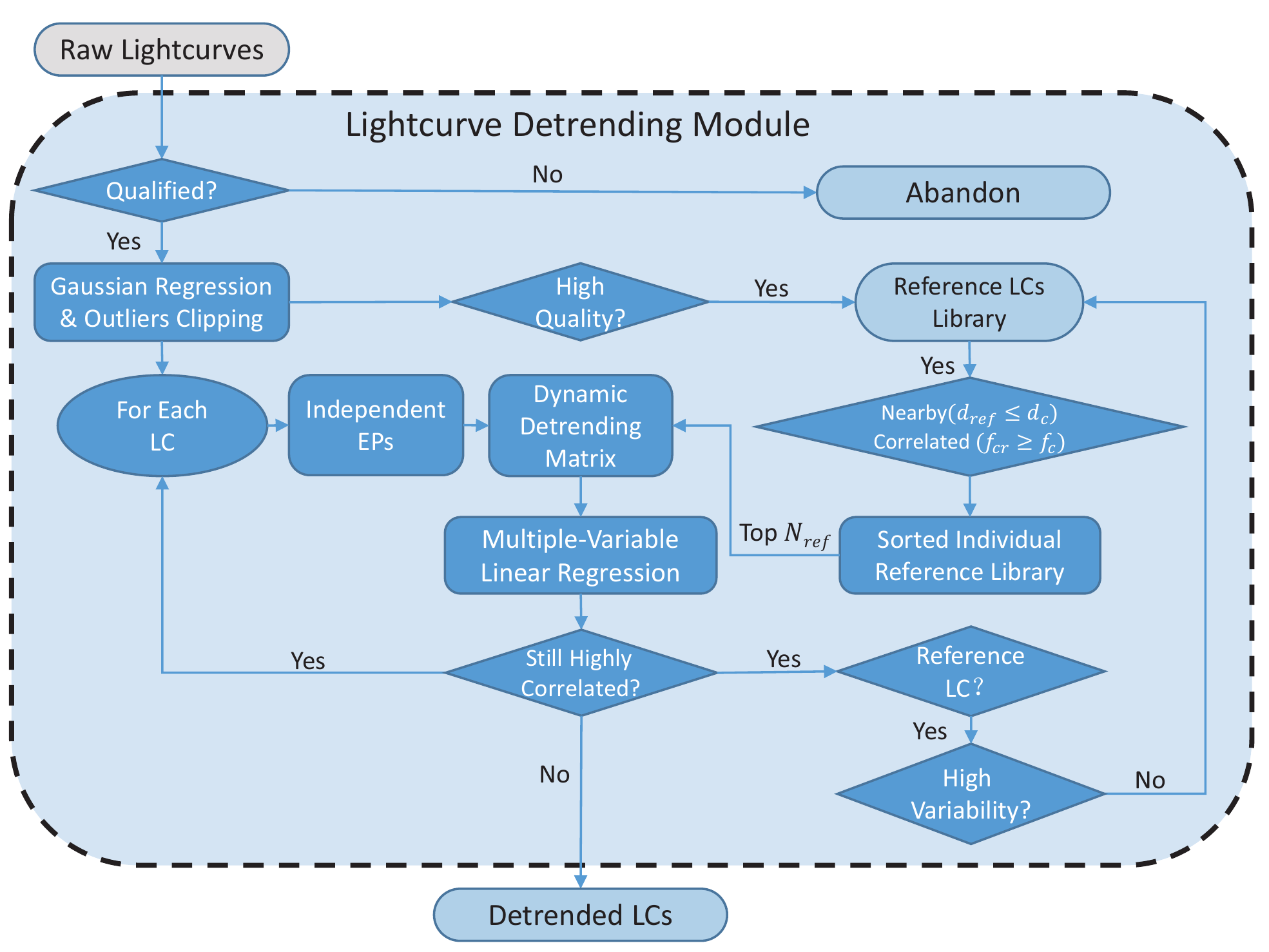}
\caption{Flow chart of our ``Lightcurve Detrending Module''.\label{fig:lc-detrend}}
\end{figure}

Wide-field photometry surveys usually suffer from serious systematic errors, e.g., scattered light, imperfect flat-fields, and tracking errors. To identify all sources of such systematics and remove them is not an easy task and sometimes is not even feasible \citep{Mekarnia16}. The widely-used simple ensemble photometry (e.g., \citealp{Honeycutt92}) assumes that all stars in the field are affected by the same systematics (e.g., transparency changes), and can be modeled by averaging out all incoming fluxes from the ensemble (or selected reference stars). However, even when color and spatial terms due to differential extinction are applied, systematic errors remain. To reveal tiny transit signals (e.g., $\lesssim 1\%$) it is essential to reach photometric precisions of a few mmag (RMS). This task is approaching the fundamental limits for ground-based photometry surveys and requires further treatments of systematics.

Modern algorithms for systematic filtering assume that while systematic errors are specific to each star, they can be modeled by a linear combination of the systematics of selected reference stars and auxiliary measurable quantities such as the centroid position and width of the point spread function \citep{Bakos07}. The determination of coefficients for this linear combination is usually performed by standard least-squares techniques. Two kinds of  method have been extensively used in the data reduction of wide-field photometric surveys: SYSREM (SYStematic effects REMove, \citealp{Tamuz05}) and TFA (Trend Filtering Algorithm, \citealp{Kovacs05}). These approaches have been proved to be effective against systematic errors from unknown sources and are also known as ``blind-detrending'' algorithms. However, sometimes they are too effective: to reach a high precision (i.e., low RMS)  lightcurves are often ``flattened out'' to the extent that there may be substantial suppression of the signals we are searching for. The success of this kind of methods is quite dependent on the selection of the reference stars. TFA (for example) uses a ``brute force'' fit of many selected template stars from the target field; if the number of template stars is too small, the detrending procedure may be ineffective, while if the number of template stars is too large, the procedure is very time-consuming and the target lightcurve may be ``over-fitted'' with all real physical variations removed \citep{Kovacs16}.

On one hand, we need to remove all unwanted signals (systematics). On the other hand, we have to retain the strength of all wanted (transit) signals. Many efforts have been made to achieve a balance between these conflicting requirements. One approach is to run the filtering methods in the ``signal-reconstructive'' mode, once the signal frequency (in the case of TFA) or basis of trends (in the case of SysRem) is determined \citep{Kovacs07}. Another approach is based on a careful selection of the template time series. \cite{Kim09} used a hierarchical clustering method to select an optimum number of co-trending lightcurves. A PCA-based criterion is used in the algorithm proposed by \cite{Petigura12} for the analysis of the Kepler lightcurves. The more involved PDC-MAP pipeline \citep{Stumpe12,Smith12} of the Kepler mission also utilizes PCA for selecting the basis vectors for the correction of systematics. In a similar manner, \cite{Roberts13} discuss the advantage of using Bayesian linear regression for robust filtering, and employing an entropy criterion for selecting the most relevant corrections.

\begin{figure}
\includegraphics[width=\textwidth]{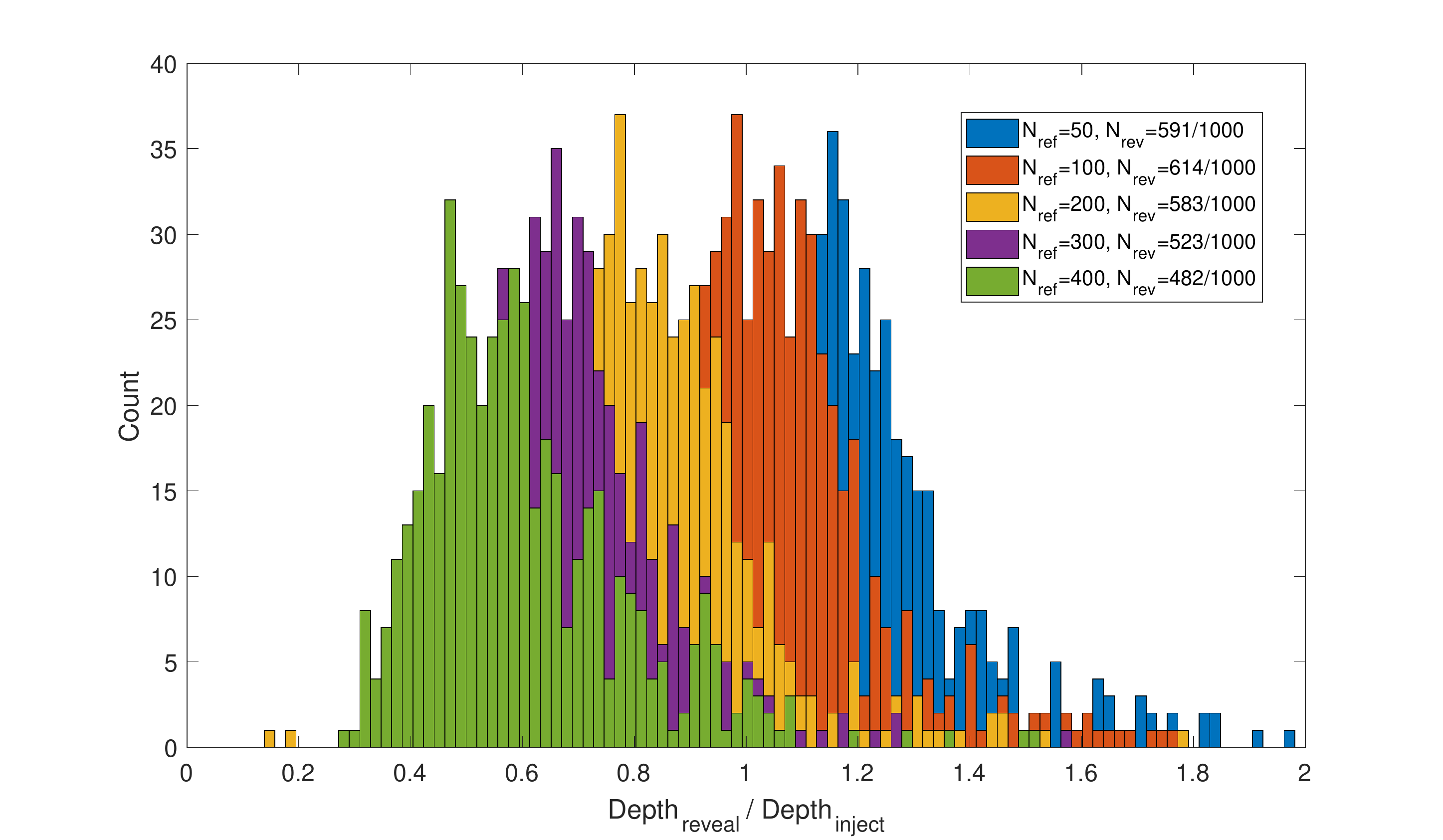}
\caption{The number of revealed transit signals versus signal distortion of the injected transits. We inject a model transit signal with a random transit depth and period into each of 1000 raw lightcurves. The total number of injected transits revealed by our pipeline is affected by $N_{\text{ref}}$, the number of reference stars within the templates. When $N_{\text{ref}}$ is too large or too small, distortion or suppression of the injected signals can be observed. The larger $N_{\text{ref}}$ is, the more serious is the suppression, so that the total revealed number of transit signals decreases with increasing $N_{\text{ref}}$. An optimum value of $N_{\text{ref}}$ for our case is around 100.\label{fig:suppression}}
\end{figure}

We implement a TFA-based algorithm with both a signal-reconstructive mode and an optimal-template selection. The detailed process is composed of the following steps (flow chart is shown in Figure \ref{fig:lc-detrend}):

\emph{Step I. filter out non-physical outliers:} these outliers, caused mainly by mis-matches of stars or bad weather, will cause serious problems in the following detrending processes. Occasionally they may also crash the BLS (Box-function Least Square Fitting algorithm \citealp{Kovacs02}) procedure when searching for transit signals. To eliminate these outliers while retaining the time-dependent astrophysical variations, the mean magnitude is subtracted from each lightcurve and a Gaussian Process Regression (GPR) model is fitted to the mean-subtracted measurements. GPR is a nonparametric kernel-based probabilistic method that can be used to predict responses of a function with multiple variables when a kernel function is given. A description of GPR models and their application in removing multi-variate systematics and intrinsic variations from lightcurve can be found in \citep{Aigrain16}. In our case, we simply set the observation time, $t_i$, as the only variate, and consider the following model of the magnitude response $m_i$:
 \begin{equation}
m_i = t_i\beta+\textit{f}_i(t),  i=1,2,3 ... N_{obs},
 \end{equation}
where $N_{obs}$ is the number of observations and $\textit{f}_i(t) \sim \text{GP}(0,\textbf{K})$, that is $\textit{f}_i(t)$ are from a zero mean Gaussian Process with covariance matrix $\textbf{K}_{ij}=k_t(t_i,t_j)$. To model the time-dependent variation, a squared exponential kernel function is adopted:
\begin{equation}
k_t(t_i,t_j) =A_t exp[-\alpha (t_i-t_j)^2].
\end{equation}
The amplitude $A_t$ and coefficients $\beta$ and $\alpha$ are estimated directly from the time series by a build-in function "fitrgp" of $\emph{\text{MATLAB}}^T$. All measurements that are $3\sigma$ away from this GPR fitted model are clipped. Note that the fitted model is not subtracted from the observations at this step.

\emph{Step II. Build a target list and a reference star library:} the target list contains all lightcurves with magnitude $\textbf{\textit{m}}_{\textit{i},apass} \leq 15.0$; the reference library contains all the stars with lightcurves having a completeness of observations greater than $90\%$ and magnitudes ranging from $\textbf{\textit{m}}_{\textit{i},apass} = 10$ to $ \textbf{\textit{m}}_{\textit{i},apass} =14$.

\emph{Step III. Build a template for each lightcurve about to be detrended:} reference stars in the individual template are selected from the reference library according to several criteria: i. the RMS of a reference lightcurve should be less than the median value of all lightcurves; ii. the angular distance between a reference star and the target should be less than $1.0^{\circ}$ to ensure they have been affected by similar kinds of systematics; iii. reference stars that are too close to the target ($\leq 30.0^{\prime\prime}$) are removed to avoid self-detrending; iv. the brightness variation of a reference star should be highly correlated with that of the target. In practice we calculate the Pearson correlation coefficient (PCC) between the target lightcurve and each valid reference lightcurve, with the top $N_{\text{ref}}$ reference lightcurves with $\text{PCC}>0.2$ being selected; v. A reference star located within the same readout channel with the target star has a higher priority to be selected. If the number of valid stars within the same readout channel is less than the required number, $N_{\text{ref}}$, we supplement the reference stars with those in nearby readout channels.

 \emph{Step IV. Build individual trend matrices for each target lightcurve:} each trend matrix contains two kind of measurements: the first $N_{\text{ref}}$ columns are the magnitudes of the $N_{\text{ref}}$ reference stars and remaining columns are external parameters of the target star, e.g., its pixel coordinates, variances of the centroid, airmass, distance to the Moon, distance to the Sun, local background variation, and FWHM and elongation of the photometric aperture. All the measurements are interpolated to the same time series as the target lightcurve.

\emph{ Step V. Multiple linear regression fit:} for each target lightcurve, we perform a Multi-variable Linear Regression on the trend matrix with each column marked as an ``independent variable'' (similarly to \citealp{Roberts13}). The resulting model is then subtracted from the target lightcurve.

\emph{Step VI. Update reference stars:} when all lightcurves have been processed, the RMS of each lightcurve is recalculated. Those reference stars with large RMS are removed from the individual reference templates and other stars with low variability are appended into the templates as reference stars.

With the updated reference stars, we then repeat Steps I--VI until the target list is empty. Lightcurves in the target list are removed if any one of the following criteria are met: i. if the RMS of a target lightcurve is almost unchanged after the last run, $\Delta \mbox{RMS} \leq 0.001$, this target lightcurve is removed from the target list since no further detrending is required on it; ii. if there is no valid reference star left in the individual template of the target star, for example, no reference lightcurve is highly correlated with the target lightcurve with $PCC>0.2$, or no reference star is located within the valid distance range, then this target is removed from the target list. Note that the loop (from step I to step VI) will be done at least once for all target lightcurves. In our experience, two loops is sufficient in most cases.

\begin{figure}
\centering
\includegraphics[width=\textwidth]{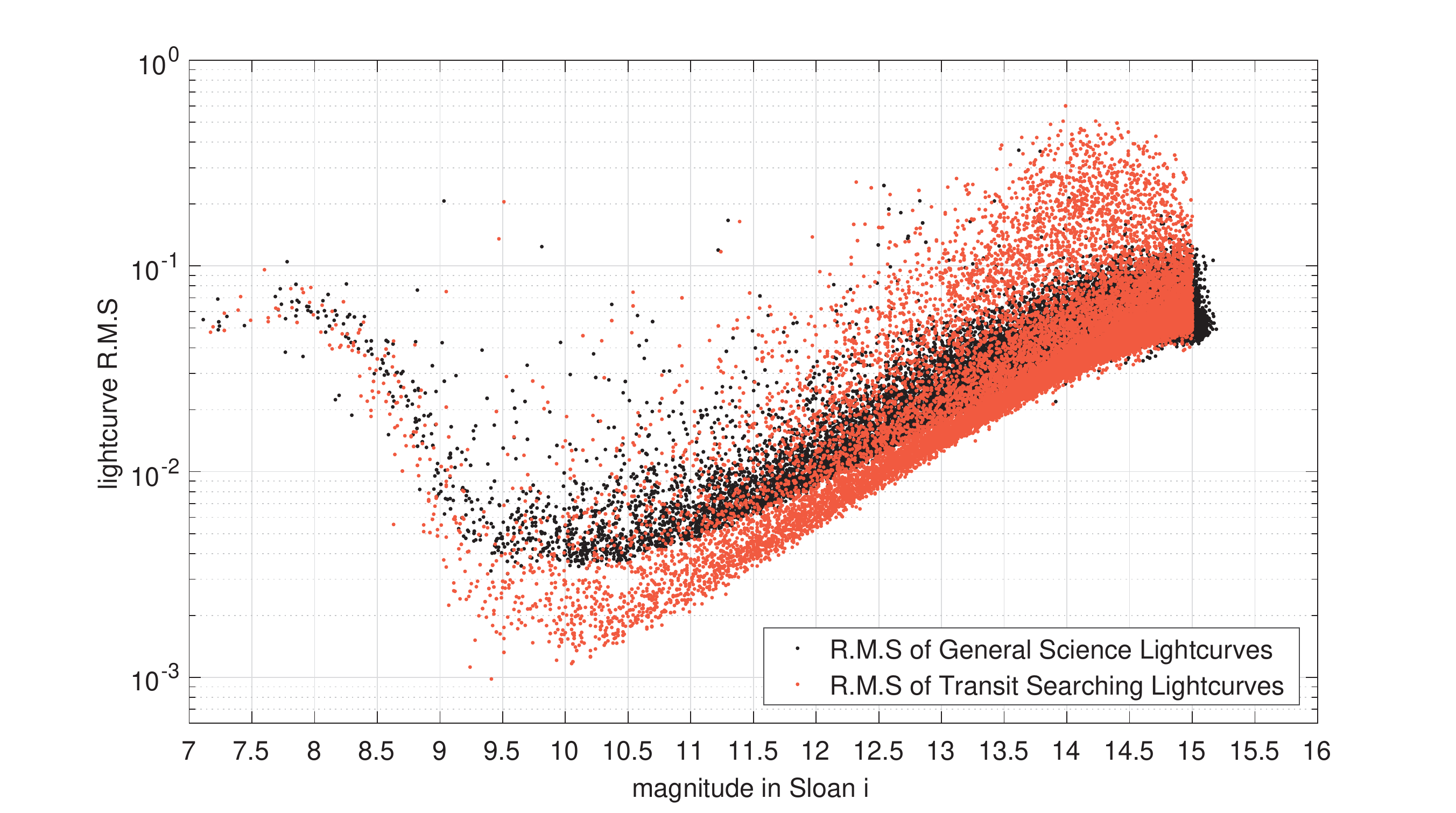}
\caption{Lightcurves RMS vs. Magnitude of 26578 stars. Each point represents the overall RMS of a detrended lightcurve with time spanning the whole observation campaign of $\sim31$ 24-hour periods. The \textbf{black} points are lightcurves with a cadence of $\sim12$ minutes and the red ones are lightcurves binned to 36 minutes. Stars brighter than $\textbf{\textit{m}}_\textit{i}=10$ mag are likely to be saturated and so suffer large variations. However, due to large extinction variations, some parts of the lightcurves of some bright stars are still unsaturated and we have found some obvious variables \citep{Zhang18} and transit candidates within this magnitude range (see Section \ref{sec:results}).\label{fig:rms}}
\end{figure}

As with any other ``blind-detrending'' algorithm, the quality of our detrending procedure depends on the number of stars in the template. Although a larger number of template stars results in a lower RMS for the lightcurve, real transit signals may be suppressed to a strength below the detection threshold of our pipeline. Furthermore, large dips caused by eclipsing binaries may be reduced in amplitude to a level where they mimic planetary transits, thereby increasing false-positive rate. To find the optimal number of template stars, we ran a series of tests on a number of lightcurves with injected transit signals. We selected 1000 raw lightcurves from our original dataset and injected modeled transit signals into each of them. The radius of the host star was fixed to $1 R_{\odot}$, while the radius and period of the transiting planet were randomly selected in the range 0.25--2.0 $R_{\text{J}}$ and 0.1--7.0 days, respectively. Raw lightcurves with injected transits were then detrended and processed by our pipeline to find transit signals. Periods of all revealed signals were compared with the injected periods to remove any false detections. Figure \ref{fig:suppression} shows the distribution of the revealed number of transits versus the signal suppression ratio, $ {\text{Depth}}_{\text{reveal}} / {\text{Depth}}_{\text{inject}}$. Five numbers of reference stars in the template were tested: $N_{\text{ref}}=50$, 100, 200, 300, 400. When the number of template stars was small, systematic errors were not removed effectively. Injected transit signals may be distorted by the remaining systematics and lead to a suppression ratio $\geq 1$. When we increase the number of template stars, the suppression ratio decreases and its median goes well below unity. The total number of revealed signals also decreases with increasing $N_{\text{ref}}$ and the optimum value is around $N_{\text{ref}}=100$. \textbf{By adopting this $N_{\text{ref}}$, we have achieved a photometric precision of $\sim$ 2.0 mmag around $m_\textit{i}=10.0$ and $\leq$ 10.0 mmag for stars brighter than $m_\textit{i}=12.5$ (see Figure \ref{fig:rms}). It is marginally enough to reveal a Jupiter-size exoplanet around a Solar-type star or an exo-Neptune around an M-dwarf. To go down below 1.0 mmag precision is quite difficult for us. The major drawbacks are the uncertainties within the flat-field correction and the intra/inter-pixel variations raised by the unperfect tracking operation that is caused by the frosting problem at Dome A. They are basically the same issue---if we could either fix the same stars on the same pixels all the time, or make a perfect flat-field, then the photometric precision will be improved significantly and allow us to detect smaller exoplanet. Unfortunately, none of them are an easy task at the Dome A, Antarctica. The good news is, as the number of sky images we acquired grows, we will be able to make better flat-fields in the future data releases.} 

\subsection{Transit Signal Searching Module}
The transit signals within a lightcurve can be modeled as a series of box-shape dips that show up periodically. One of the most effective ways to reveal this kind of signal is the BLS (Box-function Least Square Fitting) algorithm \citep{Kovacs02}. Our transit signal searching module is based on this well-tested method with some adjustments. The whole module can be divided into the following steps:

\begin{figure}
\includegraphics[width=\textwidth]{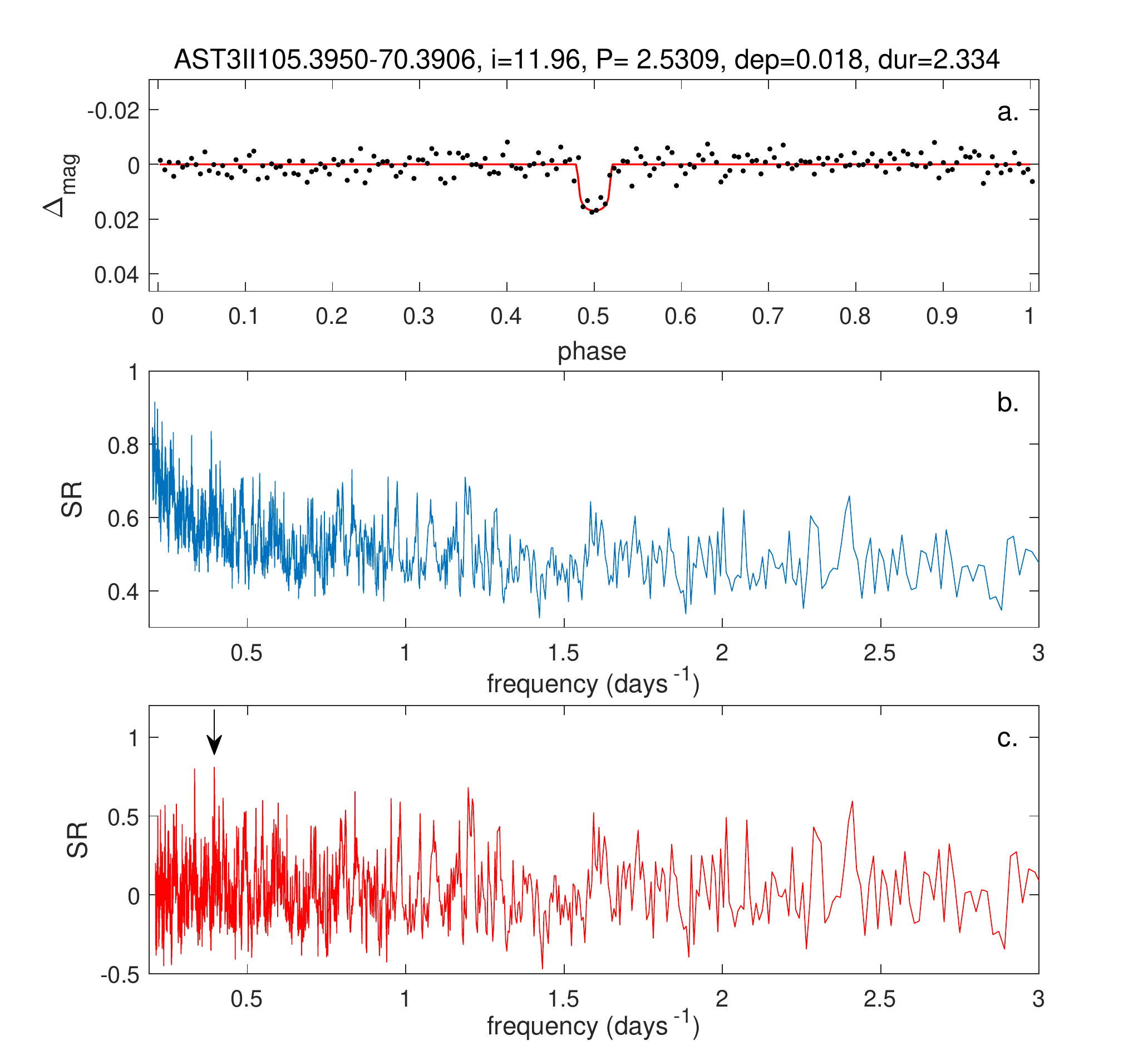}
\caption{Panel a. Phase-folded lightcurve of candidate ``AST3II105.3950$-$70.3906''; Panel b. The SR \textbf{periodogram} from the BLS algorithm; Panel c. The background subtracted SR \textbf{periodogram}. The arrow denotes the signal caused by ``AST3II105.3950$-$70.3906''.\label{fig:bls-spectrum}}
\end{figure}
\begin{enumerate}
\item \textbf{Pre-search by BLS fitting}: the purpose of the first BLS run is to generate the frequency spectrum for each lightcurve. The range of periods $P$ and the fractional transit length $q$ are set to be wide enough to enclose as many potential transit signals as possible: $0.2 \leq P \leq 10.0$ and  $0.025 \leq q \leq 0.25 $. The period resolution is set to be four times smaller than the sampling interval, which results in $\sim2000$ steps in period. At each period step, the lightcurve is folded and binned to 200 phase bins before a box-function is fitted to it. The resulting SR (Signal Residue \citealp{Kovacs02}) \textbf{periodogram} is then subtracted using a moving median filter to remove the background trend caused by low-frequency systematic errors present in the lightcurve \citep{Bakos04, Wang14}. Figure \ref{fig:bls-spectrum} shows an example of the candidate ``AST3II105.3950$-$70.3906'' whose signal is revealed after the \textbf{periodogram} detrending. We further calculate the SDE (Signal Detection Efficiency, \citealp{Alcock00, Kovacs02}) for each peak identified from this detrended spectrum. Finally, we reject all peaks with $\mbox{SDE}<1.5$ and those peaks with periods very close to integer days (e.g., $\sim 1.0\pm0.01$ days).

\item \textbf{BLS fitting over a refined period range}:  the first BLS run results in numerous candidate signals for each lightcurve. The second BLS run is to refine the parameters of signals found by the first run and filter out invalid ones. For each signal passing the first step, we perform a second BLS search in a narrow range of the earlier found period: $0.95P - 1.05P$. The period sampling number is fixed at 2000. Only the strongest signal within this period range is selected and delivered to the next step, provided the following criteria are matched:  $\& N_{\text{tr}} \geq 3$,$\Delta_{mag} \leq 0.05$ and $\mbox{SPN}\geq6.0$, where $N_{\text{tr}}$ is the number of transit dips, $\Delta_{mag}$ is the transit depth, and SPN is the Signal-to-Pink Noise ratio of the signal in the frequency spectrum. From the known statistics of exoplanets, the transit depths of confirmed planets are rarely greater than $5\%$, so it is safe to restrict the range of depths to below $0.05$; larger depths are likely to be eclipsing binaries.

\item \textbf{Further detrending by TFA with signal-reconstruction}:  as mentioned above, our lightcurve detrending module may cause some signal suppression due to over-fitting. So, when an interesting signal is revealed by the previous steps, we perform an additional detrending process to the target lightcurve with the signal-reconstructive mode. This is done by using the ``-TFA\_SR'' command in the VARTOOLS environment with the signals found by the last BLS refinement. Note that some lightcurves may contain multiple strong signals and they will be copied and detrended multiple times with the corresponding signals. This process will further reduce the RMS of the target lightcurve and enhance the strength of the target signal.

\item \textbf{Parameter filtering}:  the last step is to run BLS on each newly detrended lightcurve with a fixed period found in step 2, since the parameters of the transit signal may be changed after the TFA process with signal-reconstructive mode. The final transit signal is then filtered according to the following criteria:  $\mbox{SNR}>3$,\ \ $0.001<\Delta_{mag}<0.05$ and $0.5<T_{\text{dur}}<12$, where SNR is the signal-noise ratio of the transit signal, and $T_{\text{dur}}$ is the transit duration in hours ,\textbf{which are fitted by the BLS method}.
\end{enumerate}

\begin{figure}
\includegraphics[width=\textwidth]{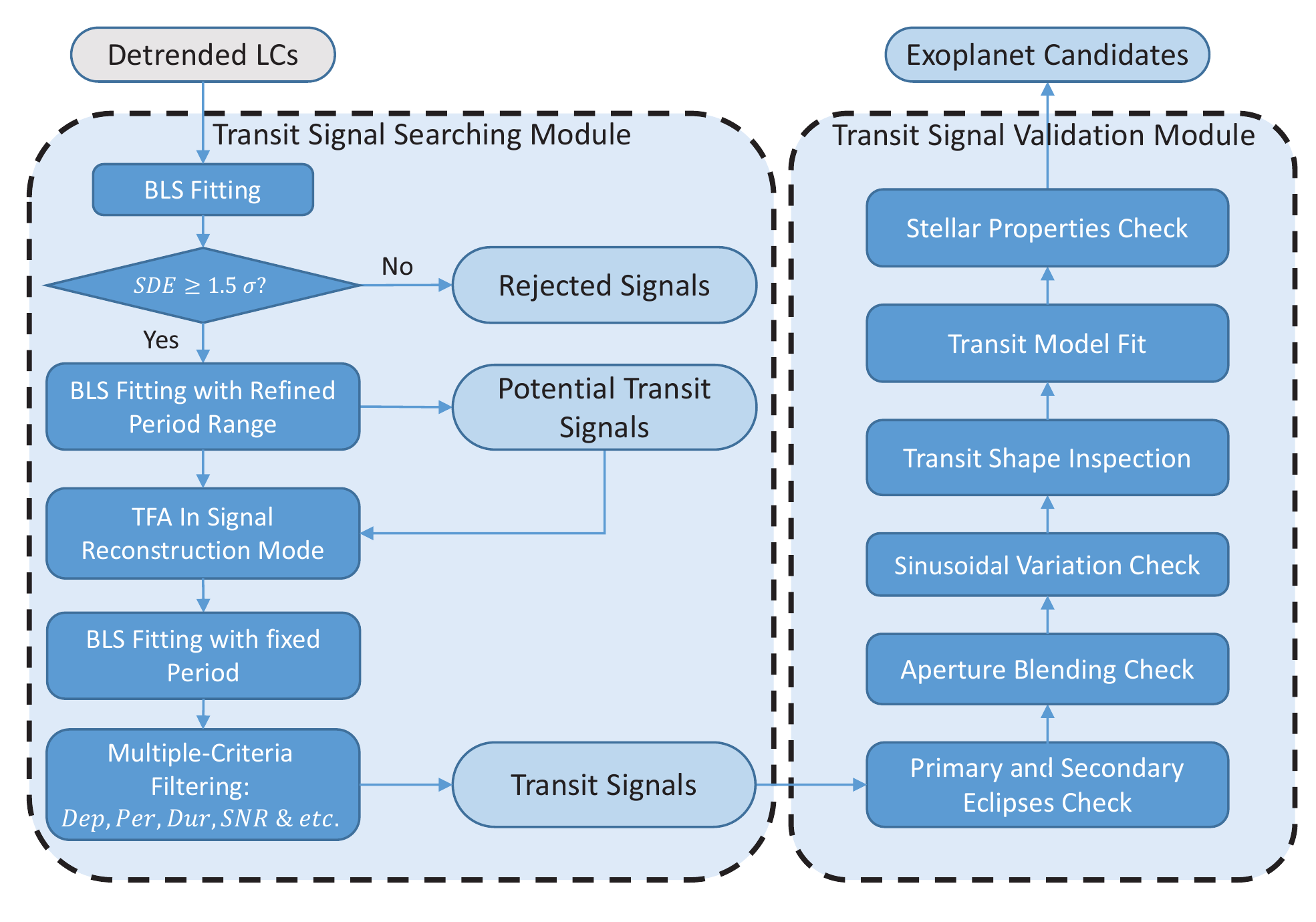}
\caption{Flow chart of the ``Transit Signal Searching Module''.\label{fig:transit-search}}
\end{figure}

After running all our data through the above steps (Figure \ref{fig:transit-search}), 1120 transit candidates (TCs) were found. This number of potential candidates is too high for easy visual inspection, so we designed a ``Transit Signal Validation Module'', described in the next section, to assist.

\subsection{Transit Signal Validation Module}
Wide-field transiting exoplanet surveys tend to suffer from a high false-positive rate, especially when the photometric precision is marginally adequate to detect transit signals. Besides systematic errors, some true astrophysical variabilities, e.g., low-mass eclipsing binaries, grazing binaries and blended background binaries, may also mimic true transit signals. Therefore, effective validation procedures must be performed before further follow-up observations. A series of validation methods have been adopted in previous successful ground-based wide-field surveys such as WASP/SuperWASP \citep{Pollacco06}, HATNET \citep{Bakos04} , HATSouth \citep{Bakos13}, KELT \citep{Pepper07}, OGLE \citep{Udalski02}, and TrES \citep{Alonso04}. Some key methods have been integrated into our validation module and proved to be effective in our previous work \citep{Wang14}. In this section we present a similar version with some adjustments for the characteristics of the AST3-II data.

Our validation module (Figure \ref{fig:transit-search}) includes quantitative analysis and visual inspections as follows:
\begin{enumerate}
\item \textbf{Primary and secondary eclipses check}: detached eclipsing binaries are one of the major sources of false alarms in many transiting exoplanet surveys. The reason is that an eclipsing low-mass star and a transiting planet will produce similar Box-shape signals, which are almost indistinguishable using the BLS method. The transit depth tells us about the secondary-to-primary radius ratio, $R_{\text{p}}/R_{\ast}$. According to the statistics of confirmed exoplanets, transiting exoplanet with transit depths $\Delta_{mag} > 0.05$ are very rare. So we may safely filter out some eclipsing binaries with very large transit depths in the previous filtering module. However, low transit depths may also be caused by a dwarf star transiting a giant or super giant host. Detecting such objects requires using other features of the light curve. Other than the transit depth, one major difference between an eclipsing binary and a transiting exoplanet is the secondary eclipse that occurs when the secondary star or the planet is blocked by the host star. Since a planet does not self-illuminate, the depth of its secondary eclipse will be much shallower than that of an eclipsing binary. So our first step is to check the existence of any secondary eclipse within a candidate lightcurve. We phase fold the lightcurve and fix the primary eclipse at phase 0.5, according to the detected period. Then we subtract the fitted model given from the BLS fitting and calculate the RMS of the residuals at phases around 0.0 and 0.5. Since the model will only fit the primary eclipse, if a secondary eclipse exists, the RMS at phase 0.0 will be significantly greater than the RMS at phase 0.5. Any candidate with $\mbox{RMS}_{\text{phase}=0.0} / \mbox{RMS}_{\text{phase}=0.5} > 2.0 $ is labeled as suspect and requires visual inspection. For each candidate that passes the above procedure, we further estimate the statistical difference between the odd and even transits, and use the significance level of the consistency in transit depth, $P_\Delta$, to determine whether the odd and even transits are drawn from the same population \citep{Wu10}. The smaller this statistic is, the more inconsistent the odd and even transits are, and the more likely the event is a false positive. The acceptance boundary is set to be 0.05 and we rejected 355 candidates with $P_\Delta < 0.05$.
\item \textbf{Aperture blending check}: a blended eclipsing binary is another major source of mimics for genuine transiting planet signals in wide-field surveys. The pixel-scale of AST3-II is $\sim 1^{\prime\prime}$ pixel$^{-1}$, so it is unlikely that more than one bright star will fall into a single pixel. However, to avoid saturating bright stars, we defocus the optics slightly to have a FWHM $\sim 5$ pixels and employ three photometric apertures: 8, 10 and 12 pixels \citep{Zhang18}. When the target field is too crowded with stars, it is likely that there will be background stellar objects within the photometric apertures of the target star. And the target star could also be contaminated by scattered light from nearby bright stars. If the contaminating star or the target star itself is in fact an eclipsing binary, its eclipsing depth will be diluted, making it look like a planetary transit. In this case, the secondary eclipse will also be made shallower, to a level that may be undetectable by our precision. So we perform a further blending check on those candidates that passed the first step. To do this we cut a stamp from the image with a size of $150\times150$ pixels for each candidate and check for blending or contaminating objects by human inspections. Since the angular resolution of AST3-II is reasonably high, this procedure is quite accurate and efficient, and 180 candidates with suspicious blending events were rejected.
\item \textbf{Sinusoidal variation check}: besides low-depth eclipses caused by binaries, some brightness variations can be caused by systematic errors or intrinsic stellar variability with a time scale similar to the planetary transit---the dimming part of the variation is easily mistaken as a dip caused by a planetary transit when we fit the phase-folded lightcurve with a box-shape function. However, measurements caused by systematics or the intrinsic stellar variability are often strongly correlated. And in such a lightcurve, the dimming and brightening parts should show up periodically, typically with a sinusoidal variation. A phase-folded lightcurve with a genuine transit event will result in only one obvious transit (dimming) detection without any strong anti-transit (brightening) detection. In this step, we calculate the ratio of improvements for the best-fit transit (dimming), $\Delta \chi_{-}^2 $, to the improvements for the best-fit anti-transit (brightening), $\Delta \chi_{+}^2 $, for each lightcurve. This measurement provides an estimate of whether a detection has the expected properties of a credible transit signal, rather than the properties of the systematic error or intrinsic stellar sinusoidal variability \citep{Burke06}. At the end of this step 320 candidates with $\Delta \chi_{-}^2 / \Delta \chi_{+}^2 < 1.5$ were rejected.
\item \textbf{Transit shape check}: a plausible transit shape is one of the most important criteria to validate a good transit candidate. Since the checks above have reduced the number of potential candidates to a level where visual inspection by human eyes is feasible, we checked each candidate independently by two authors (Dr.~Zhouyi Yu and Dr.~Ming Yang) with the same criteria: I. a complete transit dip with both the ingress and egress parts present; II. two flat ``shoulders'' before and after the transit; III. a smooth phase coverage without too many gaps. Candidates were labeled as ``bad target'' if they showed clear evidence of variability out of transit, including a secondary eclipse, an ellipsoidal variation, or a realistic variability of other forms. If both human inspectors labeled the same candidate as ``bad target'', this candidate was removed. At the end of this stage, the number of remaining candidates was reduced to 243.
\item \textbf{Transit model fit}: during this stage we perform theoretical model fits to each remaining candidate. The aim is to determine some key parameters of the transit event and filter out inconsistent systems. Before the transit model fit, we calculate the SRN (the Signal-to-Red Noise ratio) of each lightcurve. Besides the uncorrelated white noise, the errors of bright stars in ground-based wide-field photometric surveys are usually dominated by correlated red noise (Pont et al.\ 2006). Therefore, SRN is a simple and robust parameter to assess the significance of the detected transit event:
\begin{equation}
\text{SRN} =\frac{d}{\sigma_{\text{r}}}\sqrt{N_{\text{tr}}},
\end{equation}
where $d$ is the best-fitting transit depth, $\sigma_{\text{r}}$ is the uncertainty of the transit depth in the presence of red noise and $N_{\text{tr}}$ is the number of observed transit dips. The simplest way of assessing the level of red noise ($\sigma_{\text{r}}$) present in the data is to compute a sliding average of the out-of-transit data over the $n$ data points contained in a transit-length interval. This method was proposed by \cite{Pont06} and has been successfully applied to the SuperWASP candidates \citep{Christian06, Clarkson07, Kane08, Lister07,Street07}. \cite{Pont06} suggests a typical threshold range of $\mbox{SRN}\sim 7$--9 based on a red noise level of $\sigma_r \sim 3.0$ mmag. The typical value of $\sigma_{\text{r}}$ present in the AST3-II lightcurves of bright stars is $2.3$ mmag and we find that some candidates with $\mbox{SRN} \sim 5$ look plausible. To avoid missing some interesting systems, we saved all candidates with $\mbox{SRN} \geq 5$. This threshold filters out 21 candidates and passes 222 candidates.

The remaining 222 lightcurves are then modeled using the Mandel-Agol algorithm \citep{Mandel02} integrated in VARTOOLS \citep{Hartman16}. The fitted parameters of these candidates, such as period($P$), epoch, planet-to-star radius ratio ($R_{\text{p}}/R_{\ast}$), semi-major axis ($a_{\text{p}}/R_{\ast}$) and inclination ($i$) of planet's orbit are listed in Table \ref{tab:transit-fit-table}. We also calculate the ratio of the observed duration to the theoretical duration ($\eta$). This is another quantitative parameter that can be used as a filter, based on the theoretical model of the transit method. If a transit event is caused by a real planet, its transit duration, $T_{\text{dur}}$ measured directly from the phase-folded lightcurve, should be close to the theoretical duration, $T_{\text{theory}}$, calculated from the fitted parameters. This means that if $\eta\equiv T_{\text{dur}}/T_{\text{theory}} \approx 1$, the exoplanet candidate is expected to have a high probability of being a real planet. Here we employ an approximation to $T_{\text{theory}}$:
\begin{equation}
T_{\text{theory}}    = \frac{P}{\pi} \arcsin\left( \frac{ \sqrt{ (1+R_{\text{p}}/R_{\ast})^2 - (a/R_{\ast})^2 cos^2i  }}{a/R_{\ast}} \right),
\end{equation}
where $P$ is the measured period of the transit signal, $R_{\text{p}}$ is the fitted planet radius, $a$ is the fitted orbital semi-major axis, $i$ is the fitted inclination of planet's orbit, and $R_{\ast}$ is the fitted radius of the central star. This criterion was first introduced by \cite{Tingley05} in checking for candidates found by OGLE, and has been successfully applied to many WASP candidates. For each candidate, we provide $\eta$ in the 11th column of Table \ref{tab:transit-fit-table} and most of our candidates have a value close to unity.

\item \textbf{Stellar properties check}: the last step is to check the stellar radius to eliminate giant stars. For each candidate, we calculate three radii, $R_{\text{p}}^{\text{tic}}$, $R_{\text{p}}^{\text{gaia}}$ and $R_{\text{p}}^{\text{hem}}$, according to the transit depth and the stellar radii from the TIC (\textit{TESS} Input Catalog, \citealp{Stassun17}) , Gaia DR2 \citep{Gaia18} and \textit{TESS}-HERMES \citep{Sharma18} catalogs. We set a critical radius, $R_{\text{crit}} = 2 R_{\text{J}}$, to distinguish giant planets from stellar objects. If $R_{\text{p}}^{\text{tic}}\leq R_{\text{crit}}$ and $R_{\text{p}}^{\text{gaia}}\leq R_{\text{crit}}$, this candidate is labeled as a ``TC'' (Transit Candidate). If both $R_{\text{p}}^{\text{tic}}$ and $R_{\text{p}}^{\text{gaia}}$ are greater than $R_{\text{crit}} $, this candidate is removed and labeled as a ``LB'' (Low depth eclipsing Binary). We notice that many stellar radii from TIC and Gaia are not consistent, especially for dwarf stars in the TIC catalog---they are often labeled as giants in the Gaia DR2 catalog (see Figure \ref{fig:gaia-tic}). When $R_{\text{p}}^{\text{tic}}$ and $R_{\text{p}}^{\text{gaia}}$ are not consistent, e.g, $R_{\text{p}}^{\text{tic}}\leq R_{\text{crit}}$ and $R_{\text{p}}^{\text{gaia}}> R_{\text{crit}}$, this candidate is labeled by a tag of ``TC?'' which means further inspection of the stellar properties is required. If both $R_{\text{p}}^{\text{tic}}$ and $R_{\text{p}}^{\text{gaia}}$ are not available, we use $R_{\text{p}}^{\text{hem}}$ as a reference. Finally, if no stellar radius is available, these candidates are also labeled ``TC?''. \textbf{For easy retrieval, this these transit tags are listed in both Table \ref{tab:transit-fit-table} and Table \ref{tab:stellar-table}.}
\end{enumerate}

At the end of this validation module, we have 116 transiting exoplanet candidates remaining: 72 of them are strong candidates and further inspections are required for the other 44 candidates. Detailed information for all 116 candidates is listed in Table \ref{tab:stellar-table}.

\begin{figure}
\includegraphics[width=\textwidth]{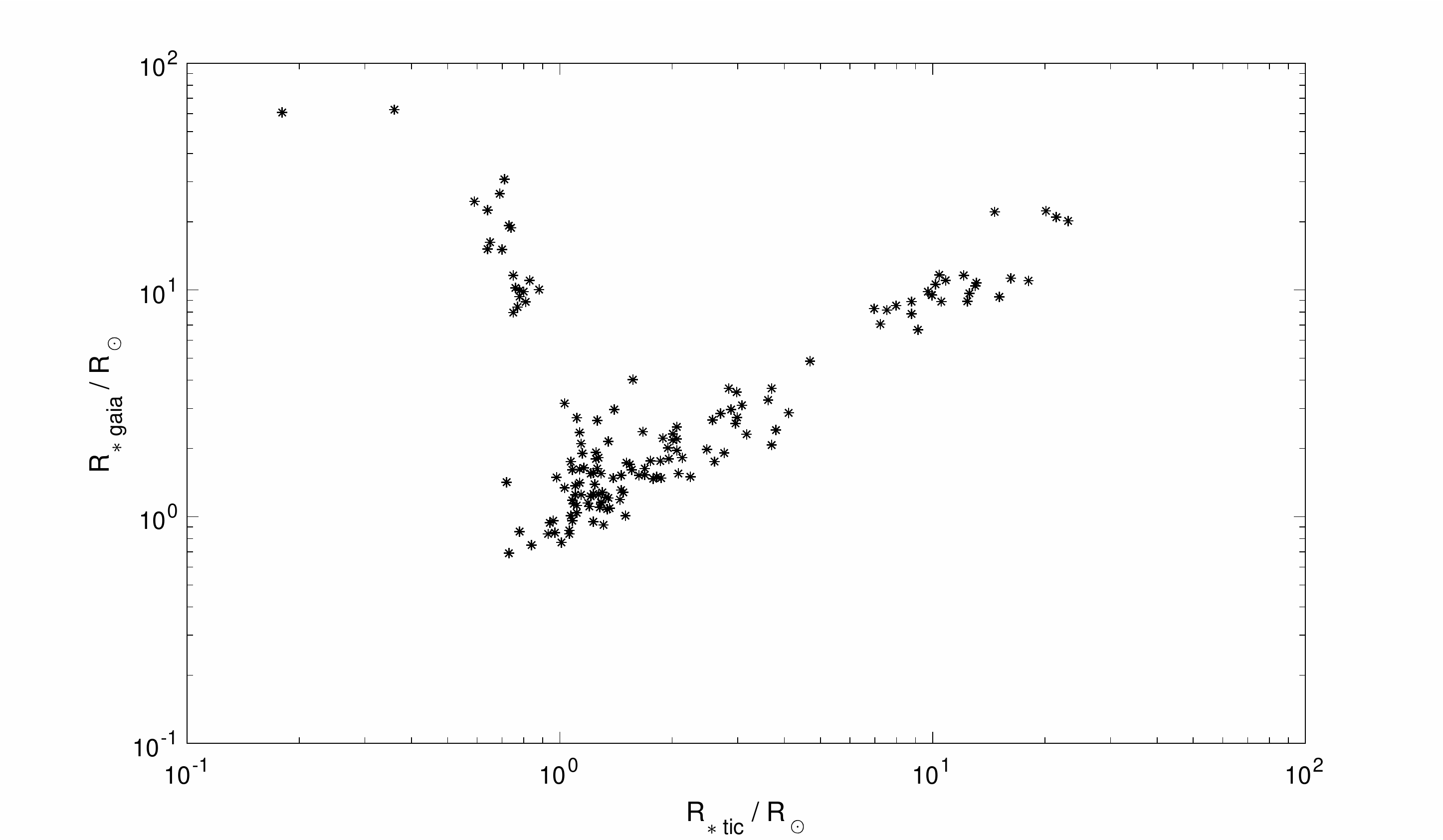}
\caption{Stellar radii of our candidates obtained from Gaia DR2 versus that obtained from TIC. Most stellar radii of our candidates are consistent between Gaia and TIC. However, many dwarf stars in the TIC catalog are labeled as giants in the Gaia DR2 database. For those targets with inconsistent stellar radii from the two databases, we label them as ``TC?'' which means that further inspection of the stellar properties are needed. \label{fig:gaia-tic}}
\end{figure}

\section{Results} \label{sec:results}

\begin{figure}
\includegraphics[width=\textwidth]{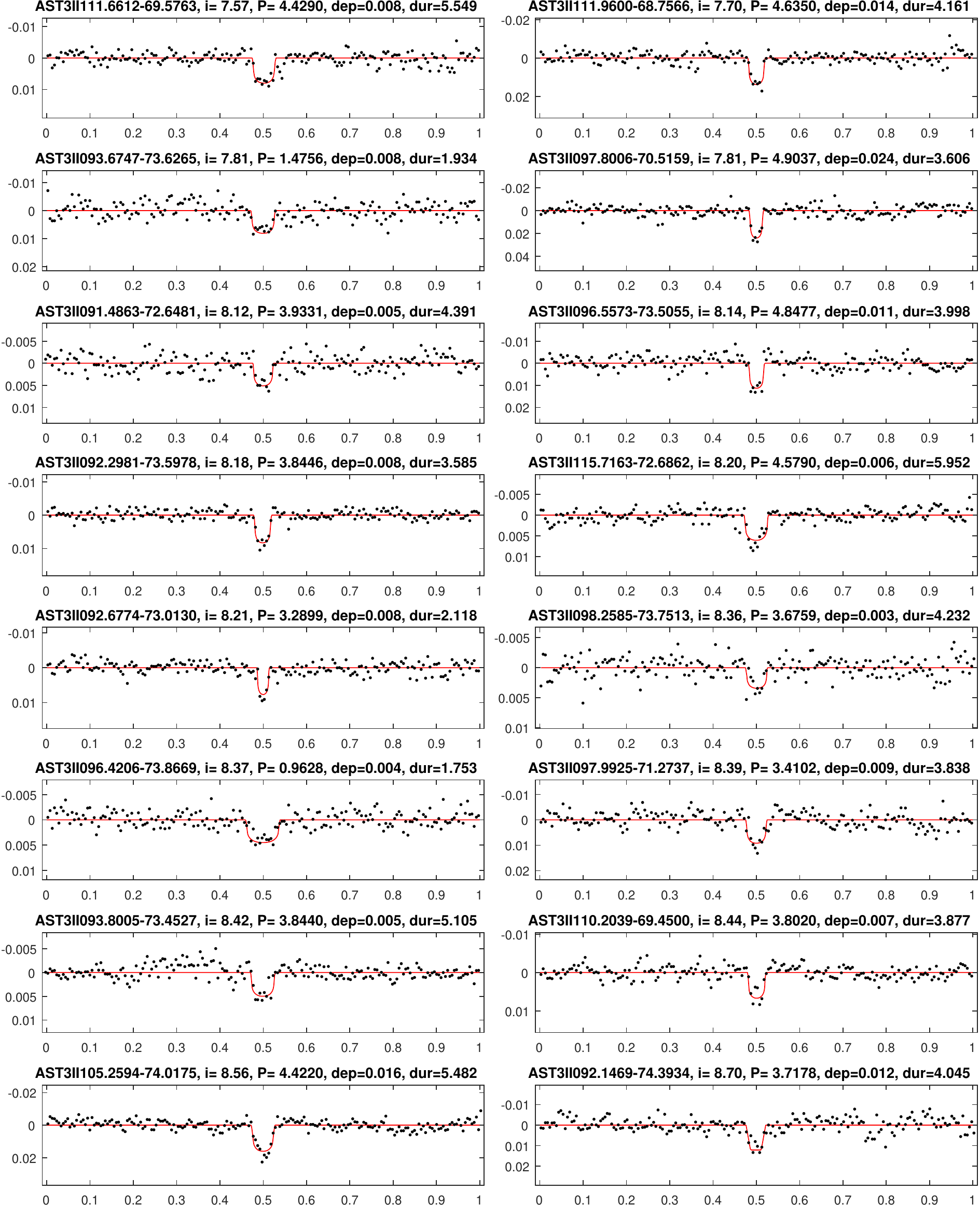}
\caption{Transiting exoplanet candidates found within the data obtained in 2016 by AST3-II. \textbf{The label above each panel contains the target ID ("AST3II+Ra+Dec"), the $\textit{i}$-band magnitude in APASS database, the period in days, the transit depth and the transit duration.} The x-axis and the y-axis of each panel are the orbital phases [0,1] and the $\Delta_{mag}$, respectively.\label{fig:transit1}}
\end{figure}

\begin{figure}
\includegraphics[width=\textwidth]{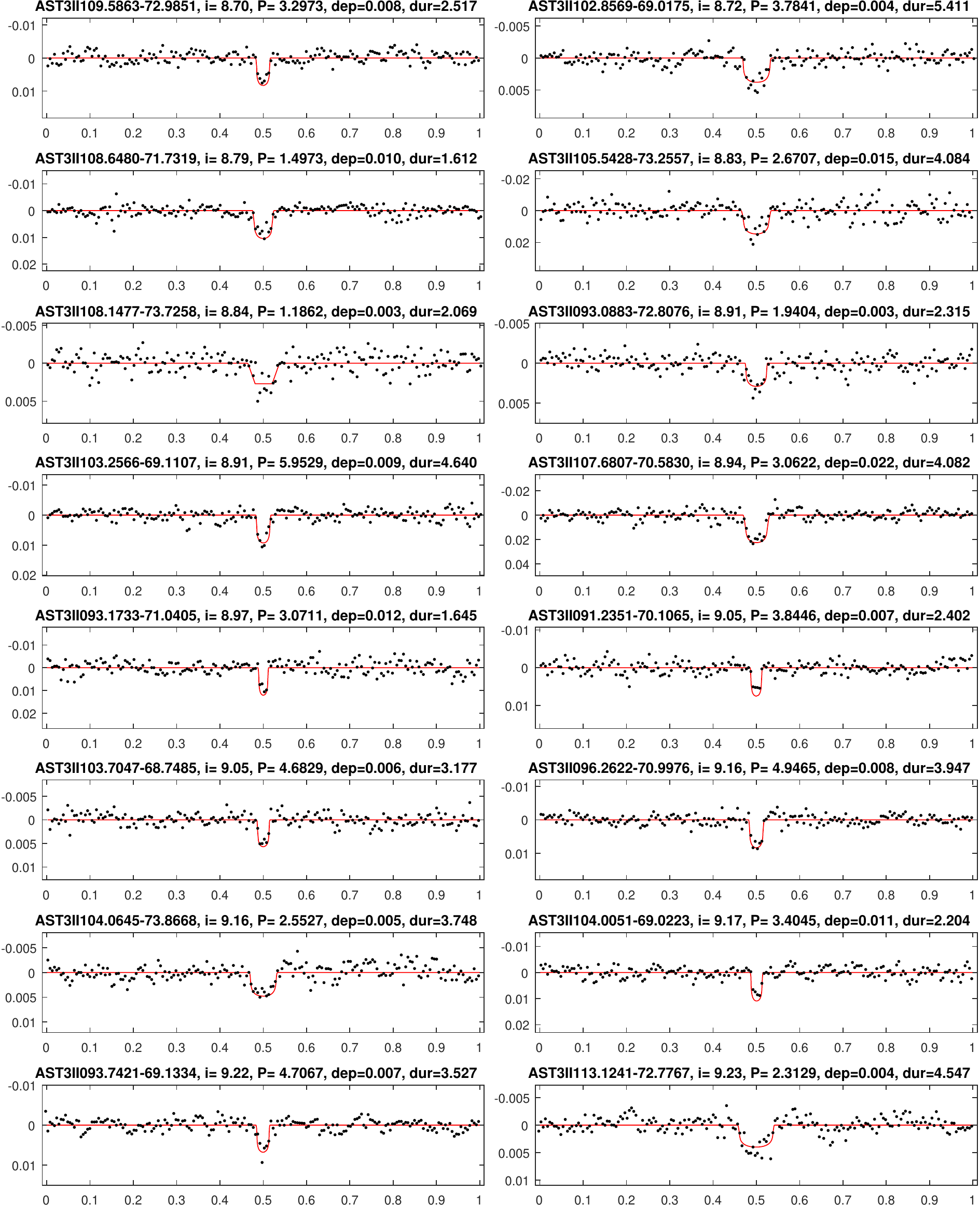}
\caption{Transiting exoplanet candidates found within the data obtained in 2016 by AST3-II, continued...\label{fig:transit2}}
\end{figure}

\begin{figure}
\includegraphics[width=\textwidth]{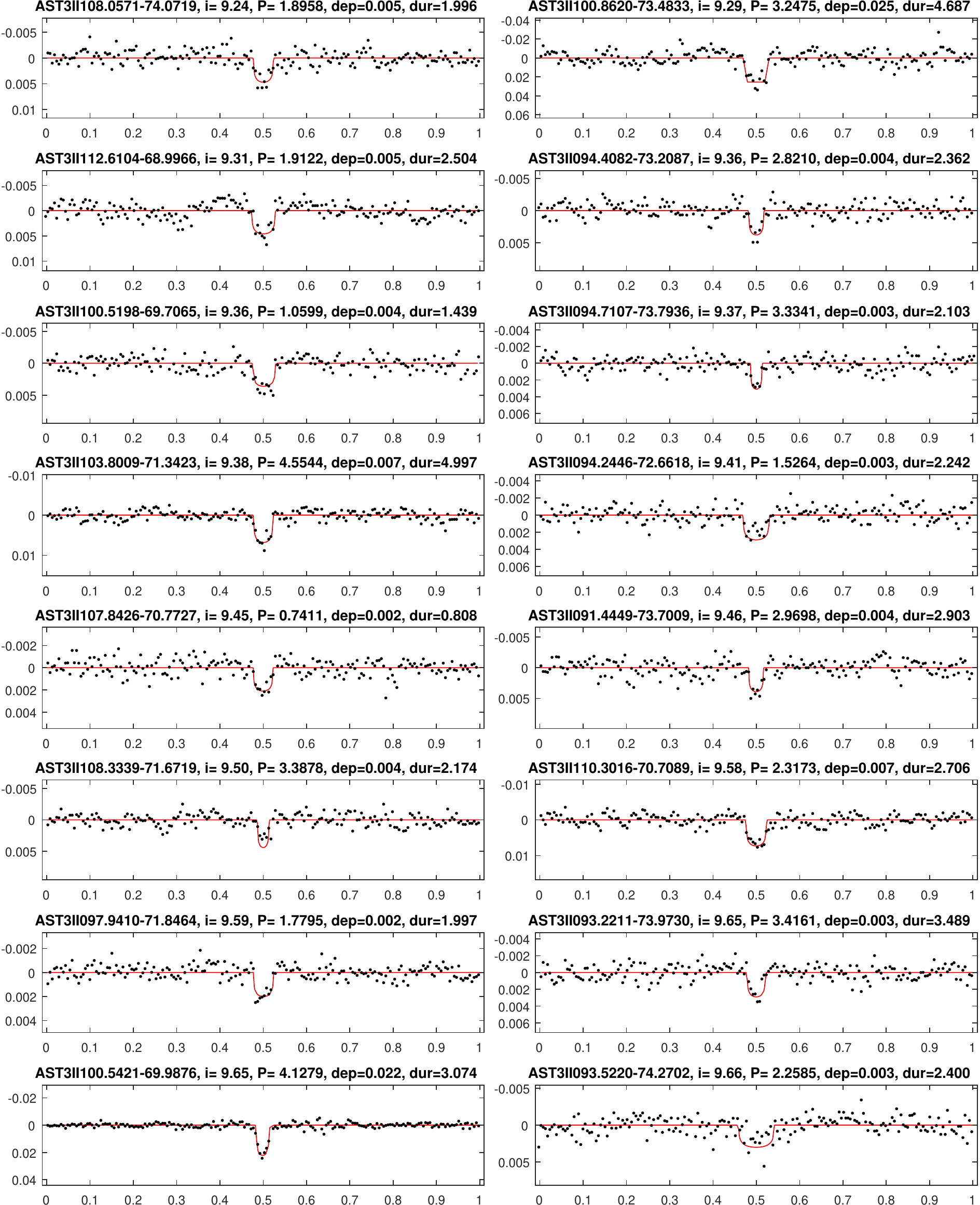}
\caption{Transiting exoplanet candidates found within the data obtained in 2016 by AST3-II, continued...\label{fig:transit3}}
\end{figure}

\begin{figure}
\includegraphics[width=\textwidth]{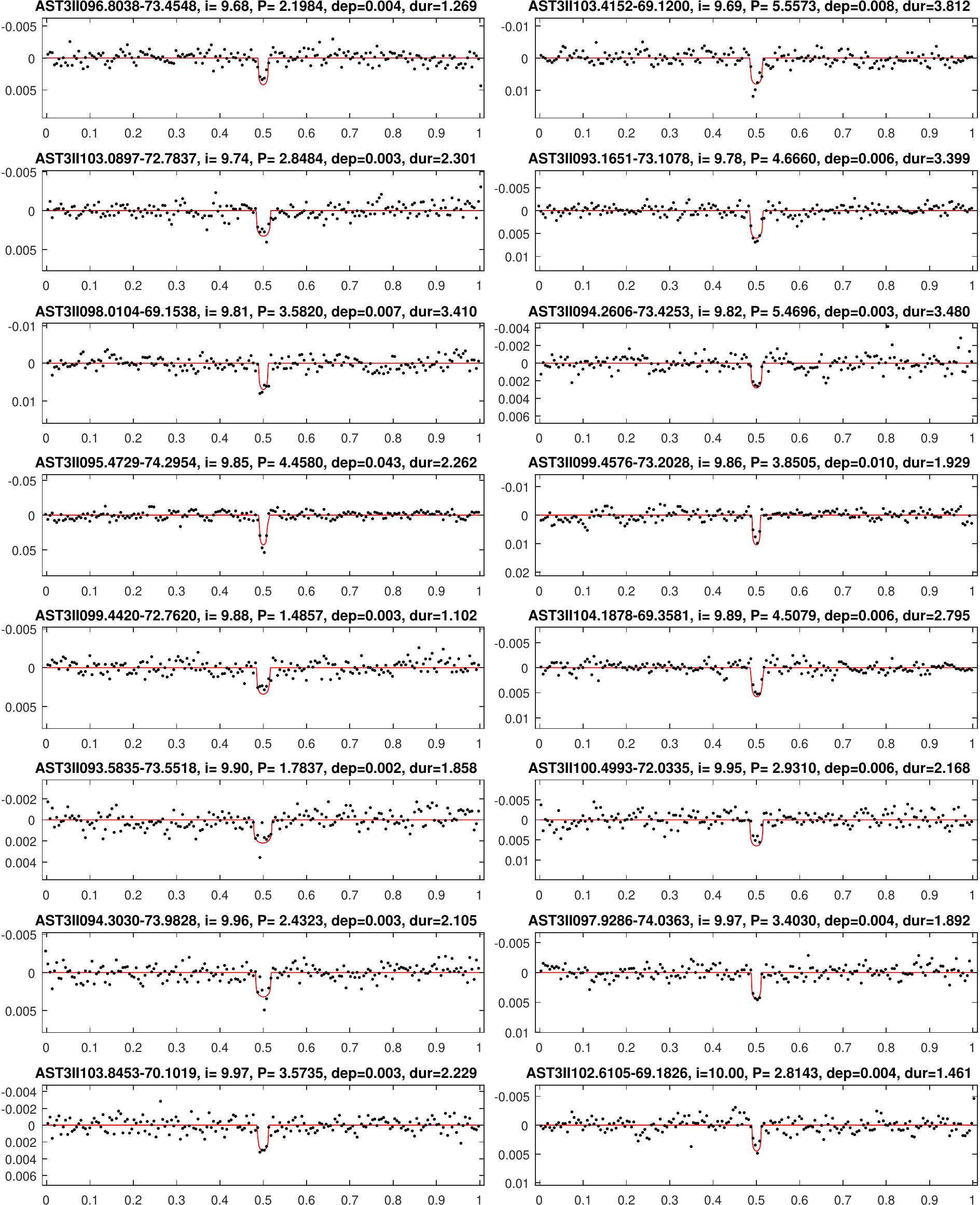}
\caption{Transiting exoplanet candidates found within the data obtained in 2016 by AST3-II, continued...\label{fig:transit4}}
\end{figure}

\begin{figure}
\includegraphics[width=\textwidth]{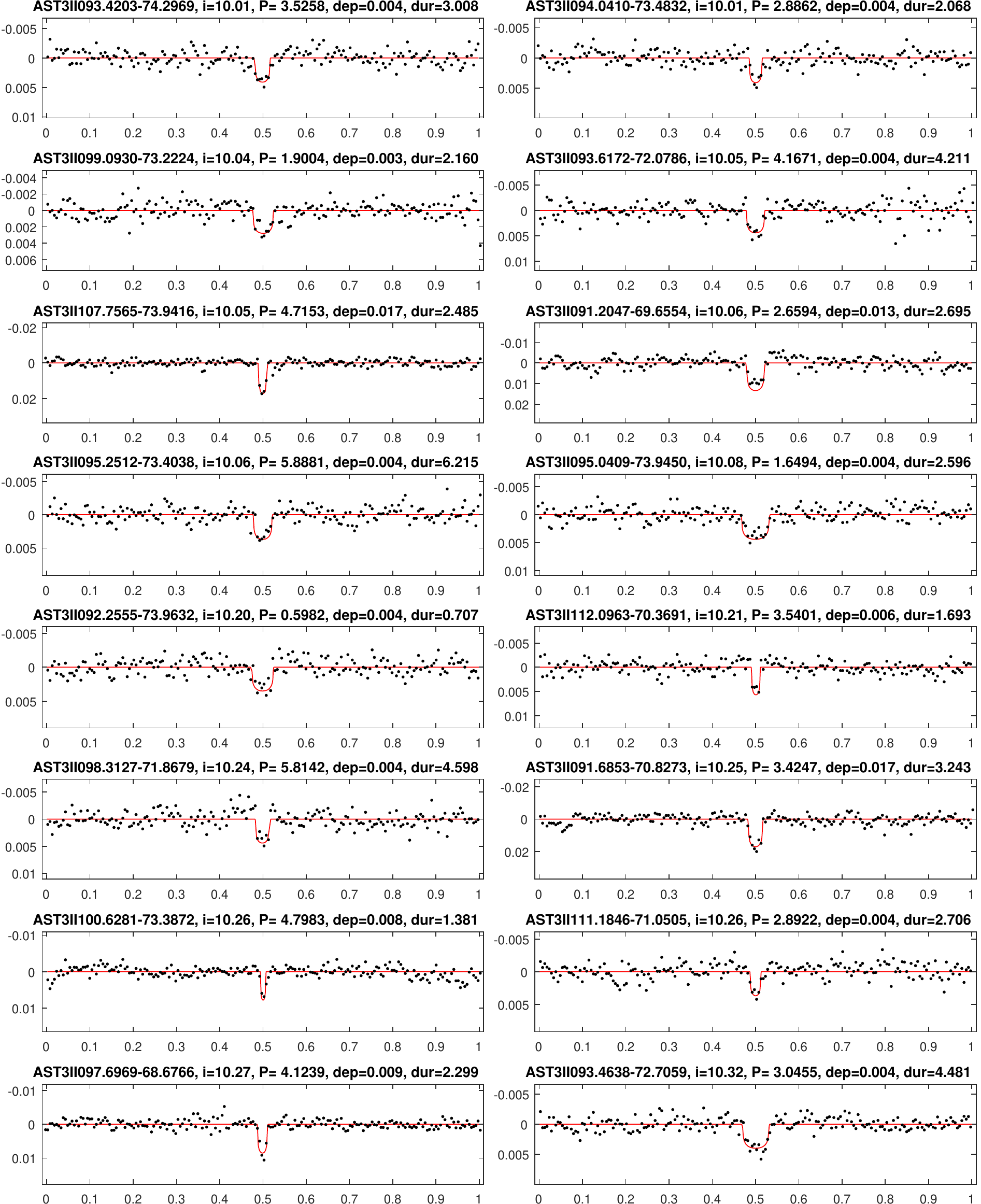}
\caption{Transiting exoplanet candidates found within the data obtained in 2016 by AST3-II, continued...\label{fig:transit5}}
\end{figure}

\begin{figure}
\includegraphics[width=\textwidth]{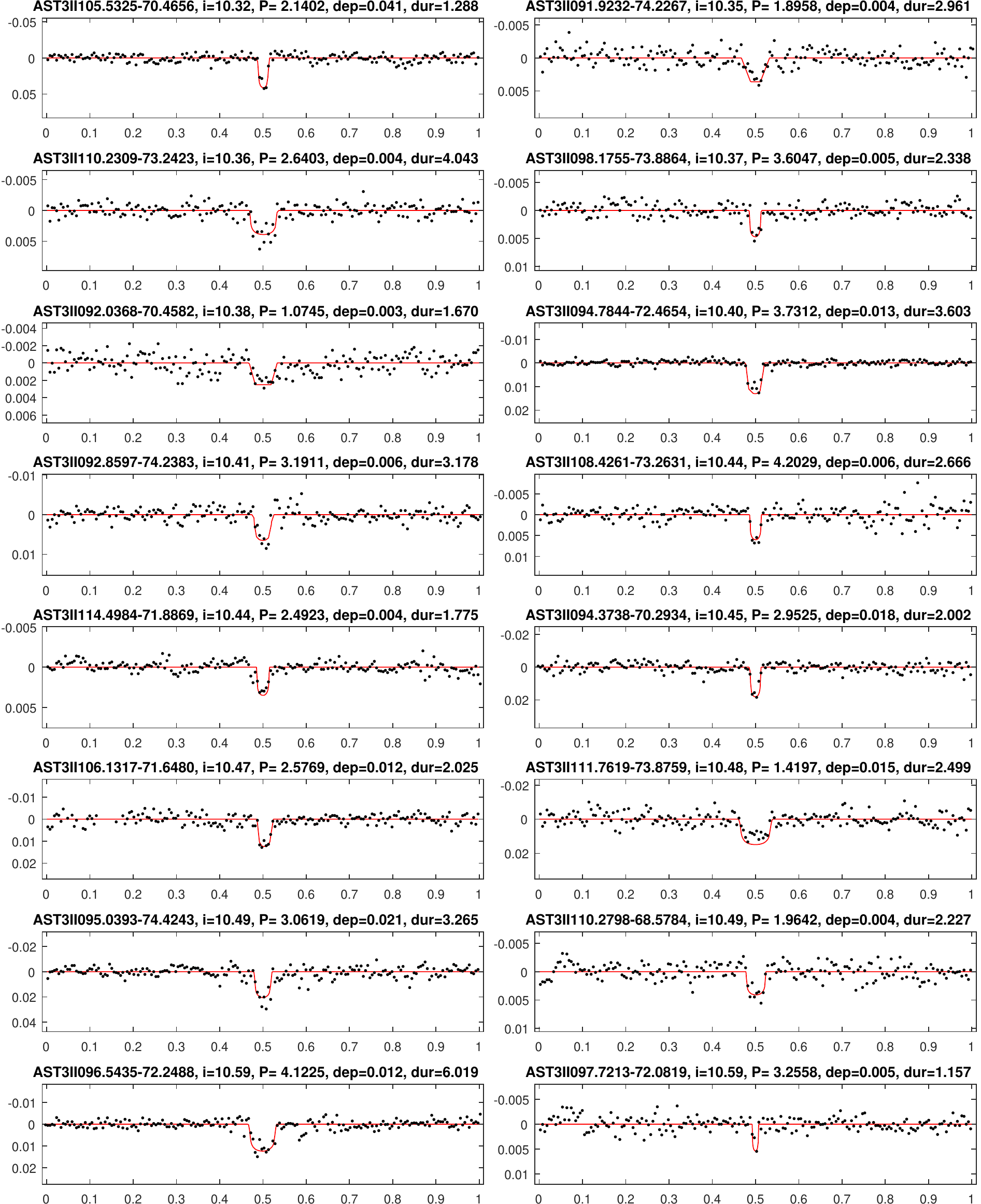}
\caption{Transiting exoplanet candidates found within the data obtained in 2016 by AST3-II, continued...\label{fig:transit6}}
\end{figure}

\begin{figure}
\includegraphics[width=\textwidth]{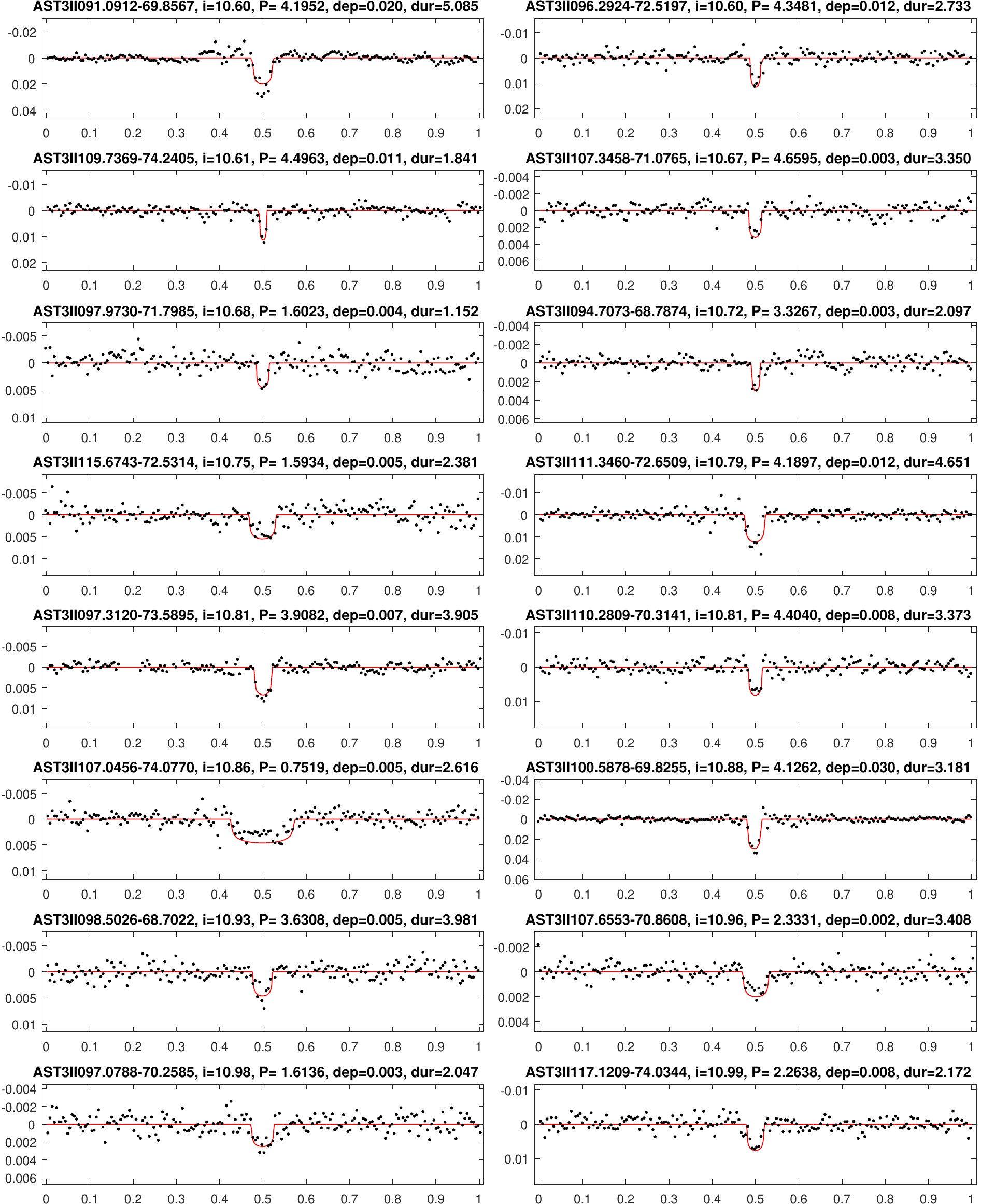}
\caption{Transiting exoplanet candidates found within the data obtained in 2016 by AST3-II, continued...\label{fig:transit7}}
\end{figure}

\begin{figure}
\includegraphics[width=\textwidth]{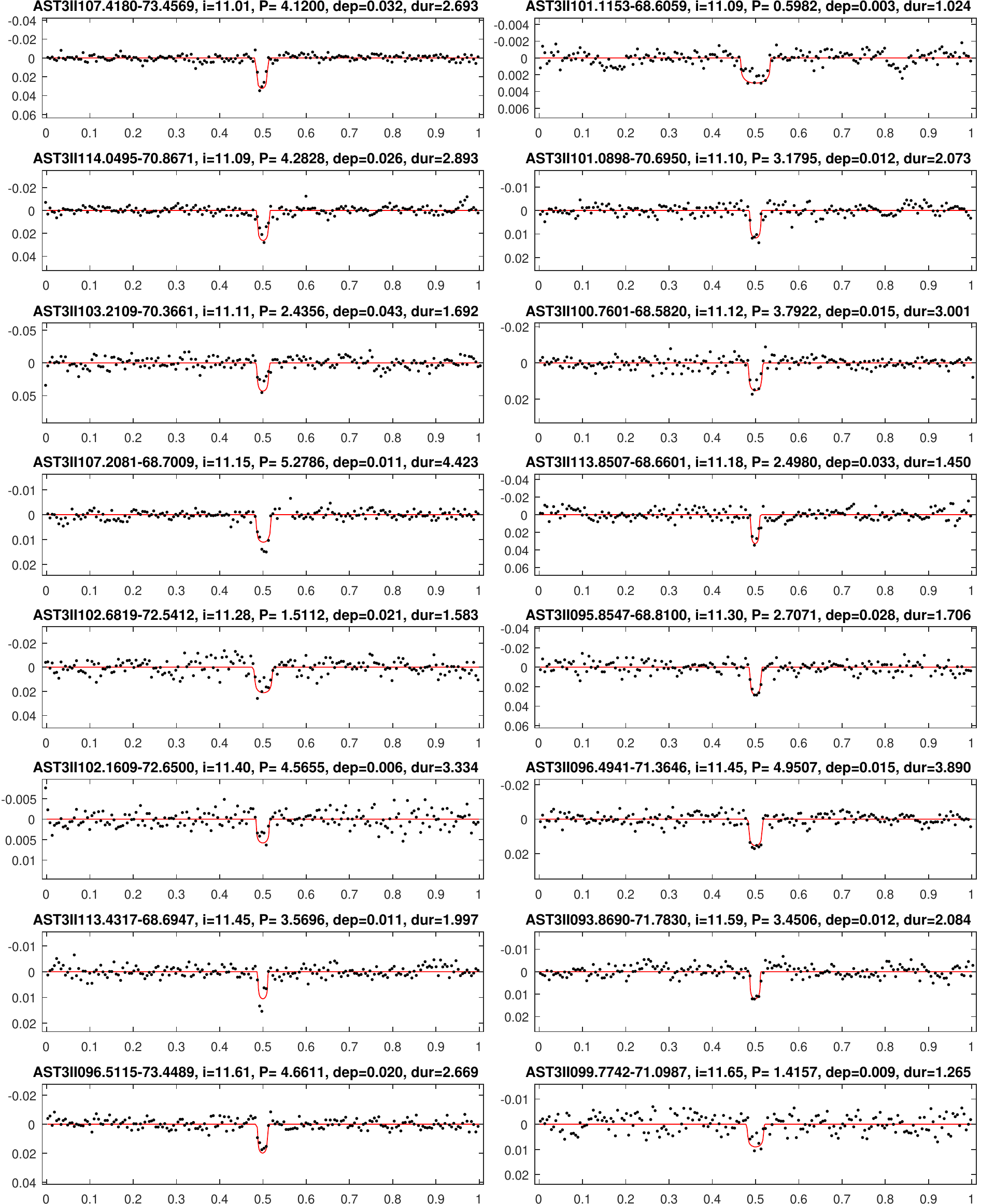}
\caption{Transiting exoplanet candidates found within the data obtained in 2016 by AST3-II, continued...\label{fig:transit8}}
\end{figure}

\begin{figure}
\includegraphics[width=\textwidth]{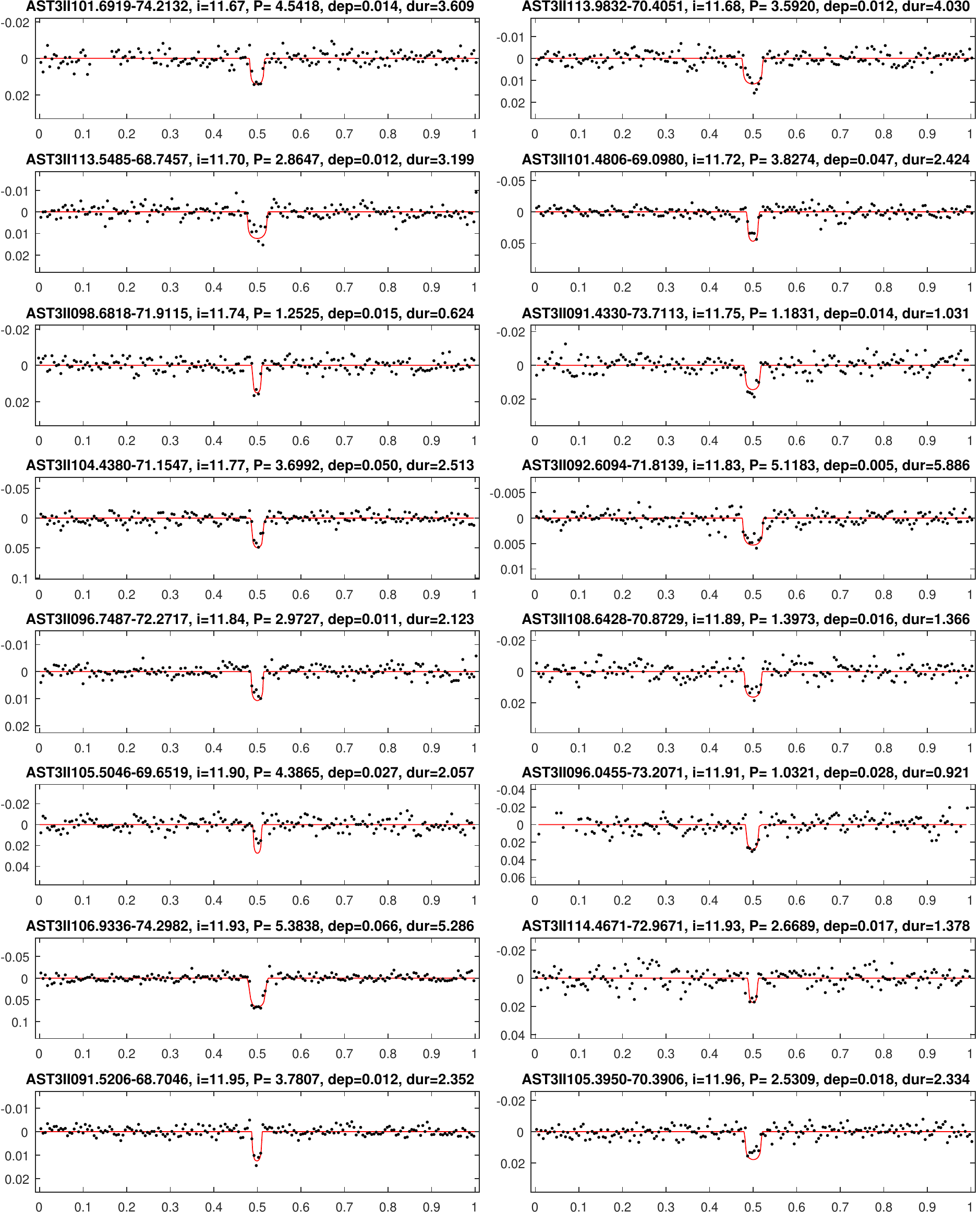}
\caption{Transiting exoplanet candidates found within the data obtained in 2016 by AST3-II, continued...\label{fig:transit9}}
\end{figure}

\begin{figure}
\includegraphics[width=\textwidth]{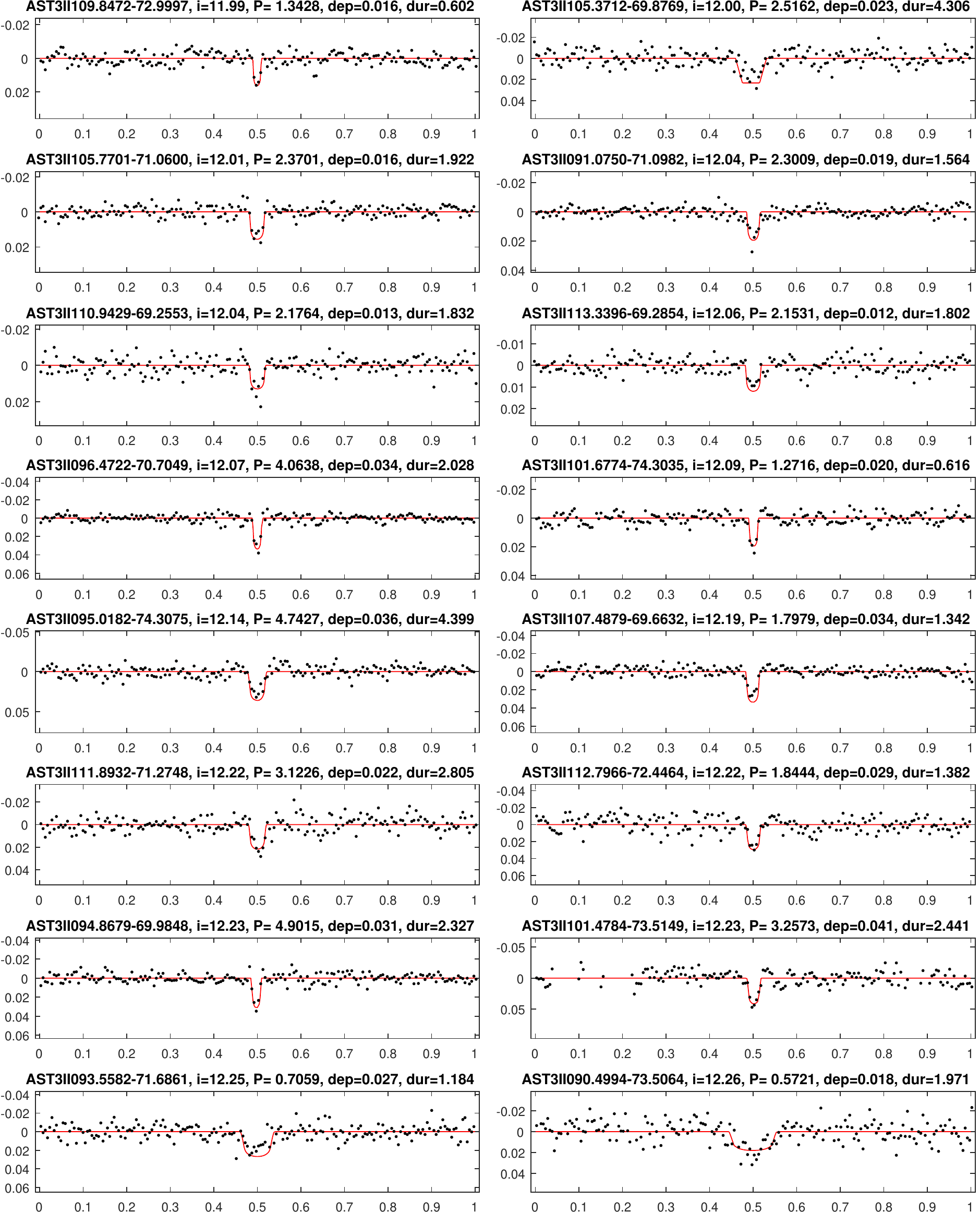}
\caption{Transiting exoplanet candidates found within the data obtained in 2016 by AST3-II, continued...\label{fig:transit10}}
\end{figure}

\begin{figure}
\includegraphics[width=\textwidth]{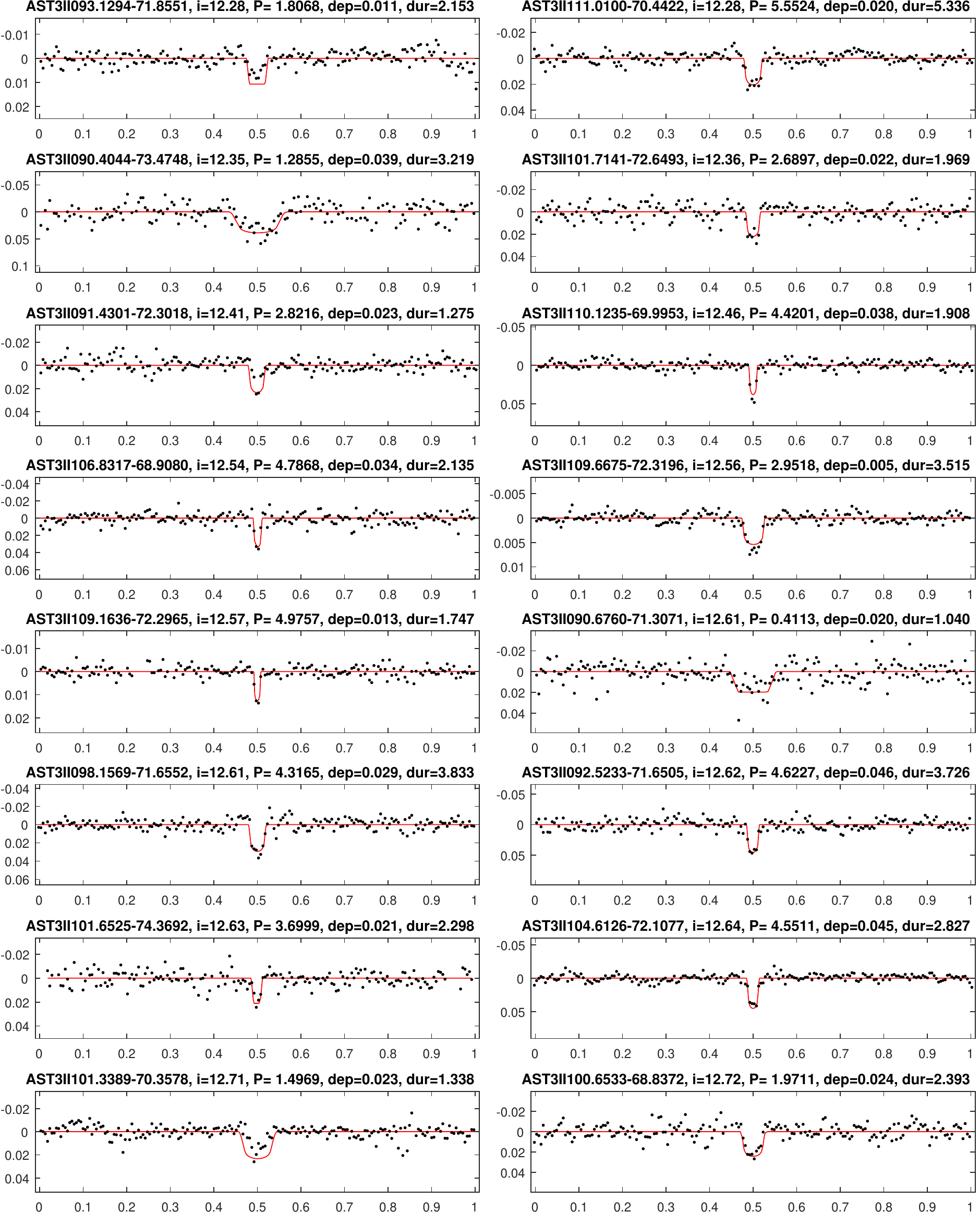}
\caption{Transiting exoplanet candidates found within the data obtained in 2016 by AST3-II, continued...\label{fig:transit11}}
\end{figure}

\begin{figure}
\includegraphics[width=\textwidth]{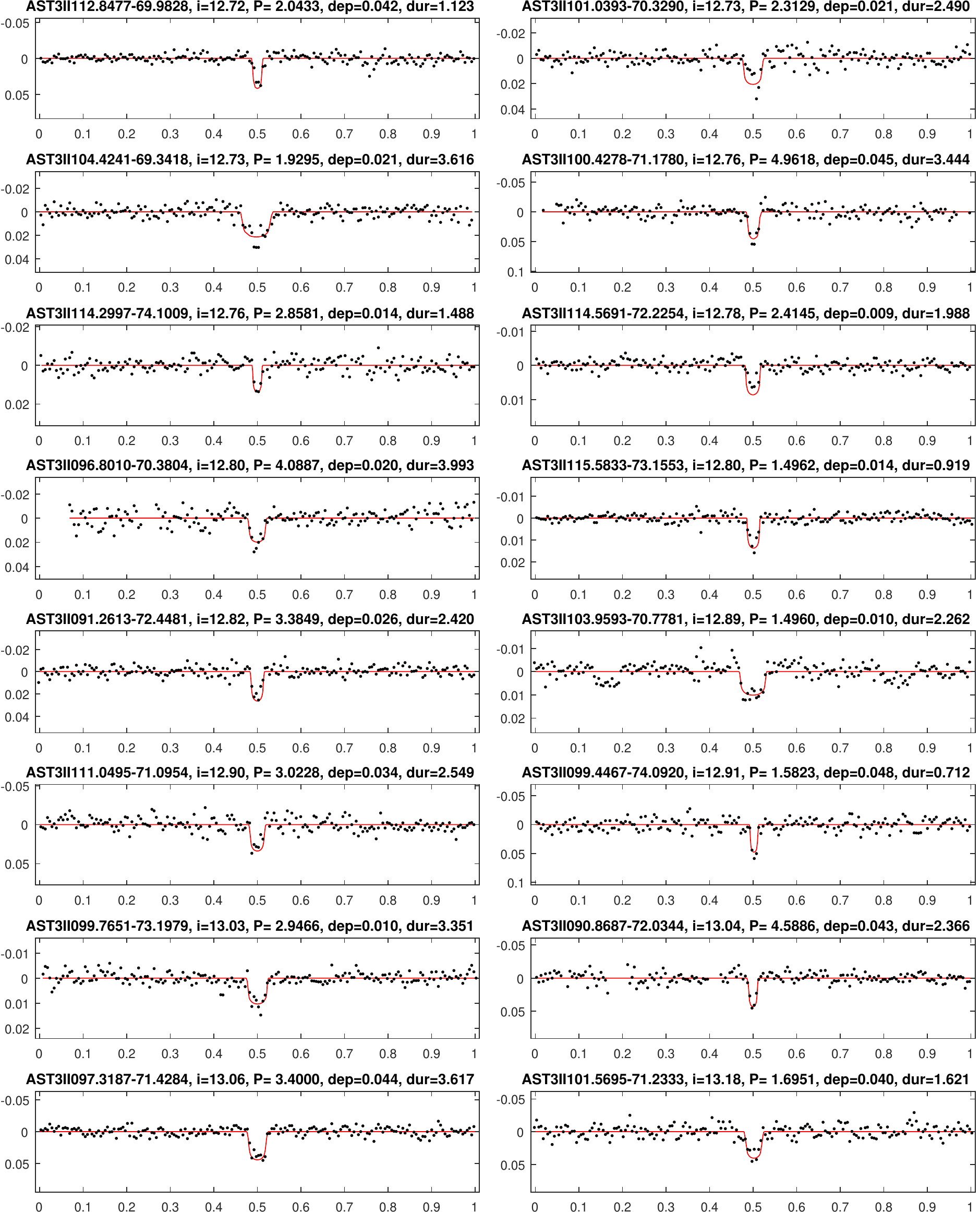}
\caption{Transiting exoplanet candidates found within the data obtained in 2016 by AST3-II, continued...\label{fig:transit12}}
\end{figure}

\begin{figure}
\includegraphics[width=\textwidth]{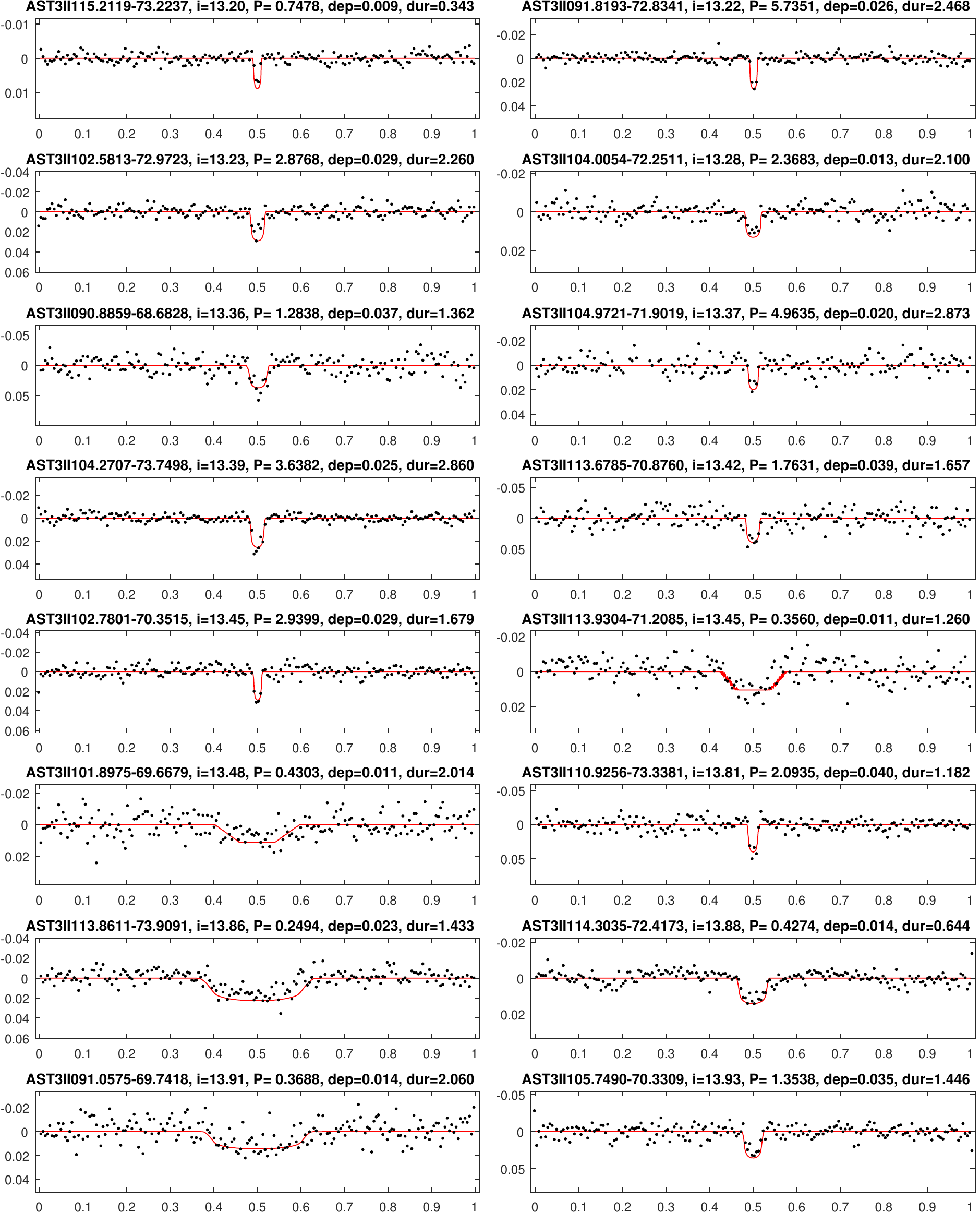}
\caption{Transiting exoplanet candidates found within the data obtained in 2016 by AST3-II, continued...\label{fig:transit13}}
\end{figure}

\begin{figure}
\includegraphics[width=\textwidth]{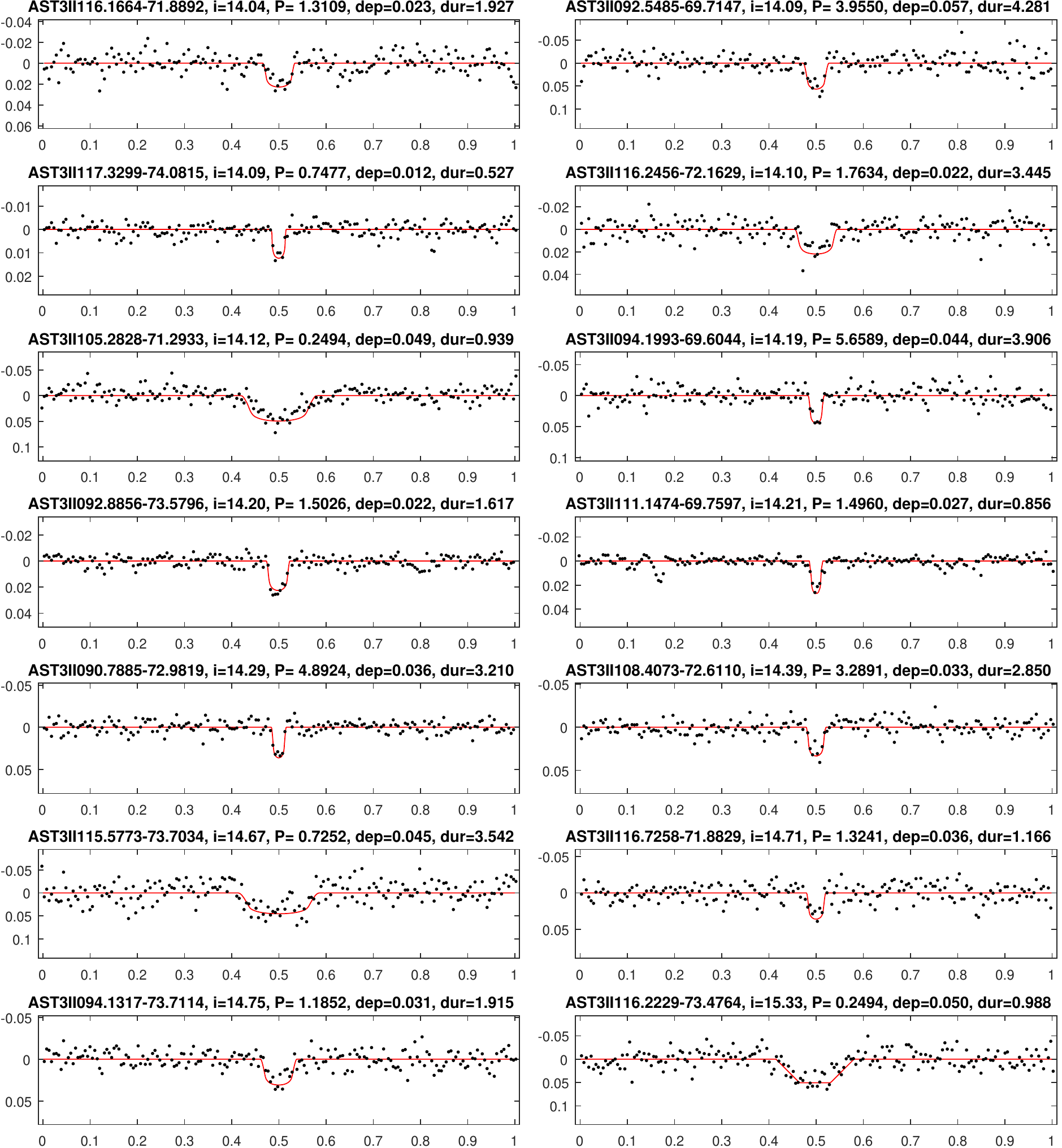}
\caption{Transiting exoplanet candidates found within the data obtained in 2016 by AST3-II, continued...\label{fig:transit14}}
\end{figure}

Our transit signal searching and validation modules have revealed a total of 222 plausible transit events. The transit signals were fitted by a Mandel-Agol model, and all the fitted parameters, including the transit epoch, period, depth, and duration, are listed in Table \ref{tab:transit-fit-table}. Lightcurves for all 222 targets are shown in Figure \ref{fig:transit1}--\ref{fig:transit14}. Each lightcurve has been folded to the fitted period and binned to 200 bins. The red solid line denotes the best-fit model for each phase-folded transit. All targets are cross-matched with the newly released TIC \citep{Stassun17}, Gaia DR2 \citep{Gaia18}, and \textit{TESS}-HERMES \citep{Sharma18} catalogs to obtain the stellar properties (such as radius) of their host stars. The planetary radius of each transit candidate is then derived according to the fitted value of $R_{\text{p}}/R_{\ast}$. The stellar properties of the host stars are listed in Table \ref{tab:stellar-table}. A tag is assigned to each target: ``TC'', ``LB'' or ``TC?'', which mean strong ``Transit Candidate'', ``Low-depth Binary'' and ``Transit Candidate but further inspections are required'', respectively. Of the 116 transiting exoplanet candidates found, 72 are strong candidates (``TC'') and 44 need further checks on their host radii (``TC?''). The smallest transit signal revealed in this work is around $\Delta_{mag} \sim 2$ mmag (e.g., target AST3II107.8426$-$70.7727, AST3II093.5835$-$73.5518 and AST3II107.6553$-$70.8608) and the longest period is around $P\leq 6.0$ days (e.g., AST3II103.2566$-$69.1107), which shows the promising capability of AST3 telescopes to find small-radius short-period transiting planets in the high-declination Antarctic sky. \textbf{Raw and detrended Lightcurves of these targets and other stars are available to the community through the website of the School of Astronomy and Space Science, Nanjing University \footnote{\url{http://www.njutido.com/tido/data.html} or  \url{http://116.62.78.33/tido/data.html}} and the Chinese Astronomical Data Center \footnote{\url{http://casdc.china-vo.org/archive/ast3/II/dr1/}}. }




\section{Summary and Conclusions}\label{sec:summary}

AST3-II is a 50 cm telescope located at Dome A (the highest point of Antarctica) enabling near-continuous observations in the \emph{i}-band during the Antarctic polar nights. It is designed to withstand the harsh climatic conditions at Dome A and has been used to perform a wide-field (FoV$\approx 1.5^\circ\times 3^\circ$) and high-resolution ($\approx 1.0^{\prime\prime}$ pixel$^{-1}$) photometric survey with a photometric precision of several millimagnitudes. Using the AST3 telescopes, the CHESPA (CHinese Exoplanet Searching Program from Antarctica) survey has been running since 2012, and 48 target fields within the Southern CVZ of \textit{TESS} were scheduled to be surveyed between 2016 and 2019.  During the austral winters of 2016 and 2017, the AST3-II telescope has successfully scanned 32 target fields.

Data from the first 10 target fields surveyed in 2016 have been fully reduced and released by \cite{Zhang18}. We have achieved a precision (RMS of the entire lightcurve) of $<2$ mmag at $\textbf{\textit{m}}_{\textit{i},\text{apass}}\approx10$ mag and $\sim50$ mmag at $\textbf{\textit{m}}_{\textit{i},\text{apass}}\approx15$ mag with a cadence of 36 minutes (Figure \ref{fig:rms}). In this work, we describe our lightcurve detrending, transit signal searching and validation modules in detail and present a catalog of 222 plausible transit signals. When combined with the stellar information given by the TIC, Gaia DR2 and \textit{TESS}-HERMES catalogs, 116 targets are labeled as transit candidates, of which 72 targets are strong candidates and 44 candidates require further inspections of the stellar parameters of their hosts. The other 106 targets are ruled out because their derived planetary radii are too large, i.e., $R_p>2 R_{\text{Jupiter}}$. Almost all of our new exoplanet candidates are listed in the input catalog of the \textit{TESS} project and some are on the high priority list (CTL). Therefore high precision photometric follow-up from \textit{TESS} will be available soon after the \textit{TESS} data release in late 2018. We are now working on obtaining radial velocity observations of our candidates to confirm them using the new {\em Veloce} facility on the 3.9m  Anglo-Australian Telescope \citep{Gilbert18}, and MINERVA-Australis facility \citep{Witt18}. The results of these follow-up observations will be presented in forthcoming papers.

\textit{TESS} was launched successfully in 2018 April and will map \textbf{most} bright stars within the southern hemisphere. Thousands of small exoplanet candidates orbiting bright nearby stars are expected to be revealed in the coming couple of years and follow-up observations with high angular resolution or different wavelengths are required. AST3-II is the first wide-field survey telescope to have worked through the long polar nights at the top of the Antarctica plateau, without any human attendance on-site during the observation campaigns. Our results demonstrate the high potential of the AST3 telescopes at Dome A to perform accurate and continuous wide-field photometric surveys. With the advantages of the polar site, the AST3 telescopes could continuously monitor hundreds of thousands of target stars around the South Ecliptic Pole for months without substantial interruption. This is particularly important for performing cross-validations of transit candidates found by  \textit{TESS}. We believe our catalog of new transit candidates within the Southern CVZ of \textit{TESS} will be a helpful reference for the flood of new candidates soon to emerge from \textit{TESS}.

\acknowledgments
This work was supported by the Natural Science Foundation of China (NSFC grants 11673011, 11333002, 11273019), National Basic Research Program (973 Program) of China (Grant Nos. 2013CB834900, 2013CB834904). The authors deeply appreciate all the CHINAREs for their great effort in installing/maintaining CSTAR, CSTAR-II, AST3-I, AST3-II and PLATO-A. This study has also been supported by the Chinese Polar Environment Comprehensive Investigation \& Assessment Program (Grant No. CHINARE2016-02-03), the Australian Antarctic Division, and the Australian National Collaborative Research Infrastructure Strategy administered by Astronomy Australia Limited. Zhang is also grateful to the High Performance Computing Center (HPCC) of Nanjing University for reducing the data used in this paper.

\software{Swarp \citep{Bertin02}, 
	VARTOOLS \citep{Hartman16}, 
	MATLAB}.

%





\appendix

\section{Appendix Data}

\begin{longrotatetable}
\centering
\begin{deluxetable}{llllllllllll}
\tablecaption{Fitted parameters of Exoplanet candidates from AST3-II observations in 2016\label{tab:transit-fit-table}}
\tablehead{
\colhead{Target ID\tablenotemark{a}} & \colhead{\emph{i}$_{\text{apass}}$\tablenotemark{b}}  &  \colhead{Epoch}    &
\colhead{$P$\tablenotemark{c}}   &\colhead{$\Delta_{mag}$\tablenotemark{d}} & \colhead{$T_{\text{dur}}$\tablenotemark{e}}   &
\colhead{SRN\tablenotemark{f}}  &  \colhead{$R_{\text{p}}/R_{\ast}$\tablenotemark{g}} & \colhead{$a/R_{\ast}$\tablenotemark{h}}   &
\colhead{$i$\tablenotemark{i}}   &   \colhead{$\eta$\tablenotemark{j}}  &   \colhead{tag\tablenotemark{k}}  \\
\colhead{AST3II$+$}  & \colhead{mag}   &  \colhead{JD2456000$+$} &
\colhead{[day]}  &  \colhead{[mag]} & \colhead{[hour]} &
\colhead{$-$} & \colhead{$-$} &  \colhead{$-$} &
\colhead{[degree]} &  \colhead{$-$} &  \colhead{$-$}
}
\startdata
AST3II111.6612$-$69.5763 &   7.57 & 1524.843179  & 4.4290   &     0.008   &    5.549   &    8.5   &   0.083    &    6.342    &    89.61    &   0.96  &   $  TC?  $ \\
AST3II111.9600$-$68.7566 &   7.70 & 1526.252284  & 4.6350   &     0.014   &    4.161   &    7.0   &   0.116    &    8.878    &    89.84    &   0.93  &   $  LB   $ \\
AST3II093.6747$-$73.6265 &   7.81 & 1513.297757  & 1.4756   &     0.008   &    1.934   &    5.9   &   0.086    &    6.088    &    89.66    &   0.96  &   $  TC?  $ \\
AST3II097.8006$-$70.5159 &   7.81 & 1513.325399  & 4.9037   &     0.024   &    3.606   &   10.2   &   0.144    &   10.860    &    89.77    &   0.91  &   $  LB   $ \\
AST3II091.4863$-$72.6481 &   8.12 & 1513.100016  & 3.9331   &     0.005   &    4.391   &    6.2   &   0.070    &    7.194    &    89.85    &   0.98  &   $  LB   $ \\
    AST3II096.5573$-$73.5055 &   8.14 & 1514.770168  & 4.8477   &     0.011   &    3.998   &    5.9   &   0.104    &    9.538    &    89.77    &   0.93  &   $  TC?  $ \\
    AST3II092.2981$-$73.5978 &   8.18 & 1513.507046  & 3.8446   &     0.008   &    3.585   &    7.6   &   0.090    &    8.556    &    89.78    &   0.96  &   $  TC?  $ \\
    AST3II115.7163$-$72.6862 &   8.20 & 1512.691684  & 4.5790   &     0.006   &    5.952   &    7.6   &   0.077    &    6.127    &    89.78    &   0.96  &   $  LB   $ \\
\enddata
\tablecomments{A machine-readable format of Table \ref{tab:transit-fit-table} is also available in the online version.}
\tablenotetext{a}{IDs of AST3-II targets in the format ``AST3II$+$RA$+$Dec'.}
\tablenotetext{b}{\emph{i}$_{apass}$: \emph{i}-band magnitudes from the APASS catalog.}
\tablenotetext{c}{$P$: the observed period of the transit signal.}
\tablenotetext{d}{$\Delta_{mag}$: the transit depth in Sloan \emph{i}-band magnitude.}
\tablenotetext{e}{$T_{\text{dur}}$: the observed transit duration.}
\tablenotetext{f}{SRN: the Signal-to-Red-Noise-Ratio of the transit event.}
\tablenotetext{g}{$R_{\text{p}}/R_{\ast}$: the planet's fitted radius as a fraction of the host star radius. The fit is based the model given by Mandel \& Agol (2002).}
\tablenotetext{h}{$a/R_{\ast}$: the fitted semi-major axis of planet's orbit in the unit of its host star radius. }
\tablenotetext{i}{$i$: The fitted orbital inclination relative to the line-of-sight of observer.}
\tablenotetext{j}{$\eta$: The ratio of the observed transit duration to the theoretical duration: $\eta = T_{\text{dur}}/T_{\text{theory}}$. }
\tablenotetext{k}{Labels from AST3-II observation: ``TC'', Transit Candidate; ``TC?'', Transit Candidate but needs further inspection; ``LB'', Low-depth Binary.}
\end{deluxetable}
\end{longrotatetable}

\begin{longrotatetable}
\centering
\begin{deluxetable}{lrclrrrrrrrlll}
\tablecaption{Stellar properties of Exoplanet candidates from AST3-II observations in 2016\label{tab:stellar-table}}
\tablehead{
\colhead{Target ID\tablenotemark{a}}   &  \colhead{ID$_{\text{tic}}$\tablenotemark{b}}  & \colhead{CTL\tablenotemark{c}} & \colhead{ID$_{\text{gaia}}$\tablenotemark{d}}   &
\colhead{$\text{mag}_i$\tablenotemark{e}} &  \colhead{$\text{mag}_{\text{tic}}$\tablenotemark{b}}  &  \colhead{$\text{mag}_{\text{rp}}$\tablenotemark{d}} &
\colhead{$R_{\ast \text{tic}}$\tablenotemark{b}}   &   \colhead{$R_{\ast\text{gaia}}$\tablenotemark{d}} & \colhead{$R_{\ast \text{her}}$\tablenotemark{f}} &
\colhead{$M_{\ast \text{tic}}$\tablenotemark{b}} &
\colhead{$P_{\text{ast3}}$\tablenotemark{g}}     &  \colhead{$R_{\text{p}}/R_{\ast}$\tablenotemark{g}} &  \colhead{tag\tablenotemark{h}}   \\
\colhead{AST3II$+$}  & \colhead{$-$}   &  \colhead{flag} &  \colhead{$-$}  &
\colhead{[mag]} & \colhead{[mag]} &  \colhead{[mag]} &
\colhead{$[R_{\odot}]$} &  \colhead{$[R_{\odot}]$} &  \colhead{$[R_{\odot}]$} &
\colhead{$[M_{\odot}]$} &
\colhead{[days]}  & \colhead{$-$} & \colhead{$-$}
}
\startdata
111.6612$-$69.5763 &  300443794 &  1      & 5267843496684819456  &  7.57 &   7.27 &    7.25  &   3.70   &   2.07   &      $-$         &  1.4177 &  4.4290 &    0.083  &   $  TC?  $ \\
111.9600$-$68.7566 &  300448730 &  0      & 5268034438043742592  &  7.70 &   7.33 &    7.37  &  15.10   &   9.32   &      $-$         &  0.7235 &  4.6350 &    0.116  &   $  LB   $ \\
093.6747$-$73.6265 &  141828072 &  0      & 5265058025713258496  &  7.81 &   6.59 &    6.57  &      $-$   &      $-$   &      $-$         &       $-$ &  1.4756 &    0.086  &   $  TC?  $ \\
097.8006$-$70.5159 &  167251167 &  0      & 5278796457859305984  &  7.81 &   6.83 &    6.81  &   4.69   &   4.85   &      $-$         &  1.2205 &  4.9037 &    0.144  &   $  LB   $ \\
091.4863$-$72.6481 &  141758352 &  0      & 5266143930882620288  &  8.12 &   7.13 &    7.29  &      $-$   &  77.39   &      $-$         &       $-$ &  3.9331 &    0.070  &   $  LB   $ \\
    096.5573$-$73.5055 &  142014846 &  0      & 5264975424897920896  &  8.14 &   7.10 &    7.23  &   0.36   &  62.38   &      $-$         &  0.3458 &  4.8477 &    0.104  &   $  TC?  $ \\
    092.2981$-$73.5978 &  141805865 &  0      & 5265113413611253888  &  8.18 &   7.27 &    7.33  &   0.64   &  15.14   &      $-$         &  0.6220 &  3.8446 &    0.090  &   $  TC?  $ \\
    115.7163$-$72.6862 &  272127358 &  0      & 5263146559102777728  &  8.20 &   7.69 &    7.76  &   9.96   &   9.48   &      $-$         &  0.7059 &  4.5790 &    0.077  &   $  LB   $ \\
\enddata
\tablecomments{A machine-readable format of Table \ref{tab:stellar-table} is also available in the online version. }
\tablenotetext{a}{IDs of AST3-II targets in the format of ``AST3II$+$RA$+$Dec''.}
\tablenotetext{b}{IDs, \textit{TESS} magnitudes, stellar radii and stellar masses from the \textit{TESS} Input Catalog \citep{Stassun17}.}
\tablenotetext{c}{CTL flags, 1 means this target is also present in the Candidate Target List of \textit{TESS} \citep{Stassun17}.}
\tablenotetext{d}{IDs, RP magnitudes and stellar radii from Gaia DR2 \citep{Gaia18} .}
\tablenotetext{e}{$i$-band magnitudes from the APASS catalog.}
\tablenotetext{f}{Stellar radii from the \textit{TESS}-HERMES survey \citep{Sharma18}.}
\tablenotetext{g}{Periods and planetary-stellar radius ratios from AST3-II observations.}
\tablenotetext{h}{Labels from AST3-II observation: ``TC'', Transit Candidate; ``TC?'', Transit Candidate but needs further inspection; ``LB'', Low-depth Binary.}
\end{deluxetable}
\end{longrotatetable}

\end{CJK*}
\end{document}